\documentclass[%
 reprint,
superscriptaddress,
 amsmath,amssymb,
 aps,
floatfix,
]{revtex4-2}
\usepackage{xcolor}
\usepackage{graphicx}
\usepackage{graphics}
\usepackage{amssymb}
\usepackage{amsmath}
\usepackage{epsfig}
\usepackage{latexsym}
\usepackage{color}
\usepackage{rotating}
\usepackage{subfigure}
\usepackage{hyperref}
\usepackage{times}
\usepackage[capitalise]{cleveref}
\usepackage[flushleft]{threeparttable}
\usepackage{comment}
\usepackage{soul}
\usepackage{url}

\newcommand{\beq}{\begin{equation}}
\newcommand{\eeq}{\end{equation}}
\newcommand{\bea}{\begin{eqnarray}}
\newcommand{\eea}{\end{eqnarray}}
\newcommand{\bean}{\begin{eqnarray*}}
\newcommand{\eean}{\end{eqnarray*}}
\makeatother

\begin{document}

\title{Massive quantum superpositions using magneto-mechanics}

\author{Sarath Raman Nair}
\thanks{Joint first authors}
\email[Corresponding author : ]{sarath.raman-nair@mq.edu.au}
\affiliation{School of Mathematical and Physical Sciences, Macquarie University, NSW 2109, Australia}
\affiliation{ARC Centre of Excellence for Engineered Quantum Systems (EQUS), Macquarie University, NSW 2109, Australia}

\author{Shilu Tian}
\thanks{Joint first authors}
\email[Corresponding author : ]{ shilu.tian@oist.jp}
\affiliation{Quantum Machines Unit, Okinawa Institute of Science and Technology Graduate University, Onna, Okinawa 904-0495, Japan}

\author{Gavin K. Brennen}
\affiliation{School of Mathematical and Physical Sciences, Macquarie University, NSW 2109, Australia}
\affiliation{ARC Centre of Excellence for Engineered Quantum Systems (EQUS), Macquarie University, NSW 2109, Australia}

\author{Sougato Bose}
\affiliation {Department of Physics and Astronomy, University College London, Gower Street, WC1E 6BT London, UK}

\author{Jason Twamley}
\affiliation{Quantum Machines Unit, Okinawa Institute of Science and Technology Graduate University, Onna, Okinawa 904-0495, Japan}

\vspace{10pt}

\begin{abstract}
Macroscopic quantum superpositions of massive objects are deeply interesting as they have several potential applications ranging from the exploration of the interaction of gravity with quantum mechanics to quantum sensing.
In this article, we propose two related schemes to prepare a spatial superposition of massive quantum oscillator systems with high Q-factor via a superposition of magnetic forces.
In the first method, we propose a large spatial superposition of a levitated spherical magnet generated via magnetic forces applied by adjacent flux qubits. 
We find that in this method the spatial superposition extent ($\Delta z$) is independent of the size of the particle.
{\color{black} In the second method, we propose a large spatial superposition of a magnetically levitated 
flux qubit, generated via driving the levitated qubit inductively. }
In both schemes, we show that ultra-large superpositions
{\color{black}$\Delta z/\delta z_{\rm zpm}\sim 10^4 \text{--}10^6$,} are possible, where $\delta z_{\rm zpm}$ is the zero point motional extent.
\end{abstract}

\maketitle

\section{Introduction\label{sec:introduction}}

{\color{black}The generation of macroscopic quantum superpositions of coherent states, sometimes known as Schrodinger Cat states }\cite{Schrodinger1935DieQuantenmechanik}, is one of the most sought-after goals in quantum mechanics. {\color{black} Such superpositions can be used to probe the domain of validity of quantum mechanics, whether gravity can entangle objects ({\color{black}i.e}. quantum gravity induced entanglement of masses, QGEM),
the validity of various collapse models \cite{Penrose1996OnReduction, Bassi2013ModelsTests, Vinante2017ImprovedCantilevers, Helou2017LISAModels, Zheng2020RoomMicro-oscillator}, and for extreme high precision sensing \cite{Rademacher2020}.} {\color{black}By preparing multiple macroscopic quantum superposition states, one can test the ability of space-time itself to exist in a quantum superposition state \cite{bose2016matter,Bose2017,Marletto2017,Bhole2020WitnessesSystems},} and this topic has attracted much attention and discussion recently \cite{marshman2020locality,torovs2020relative,bose2022mechanism,Altamirano2018GravityChannel, Hall2018OnGravity, Belenchia2019InformationSuperposition, Carlesso2019TestingSuperposition, Christodoulou2019OnGeometries, Nguyen2020EntanglementInteraction, Bhole2020WitnessesSystems, Krisnanda2020ObservableGravity}. 
It would be highly desirable to make Schrodinger cats that have both large mass and spatial extent {\color{black} as such states would be more useful for the above mentioned applications}.

{\color{black}Though there are many proposals for the generation of spatial superpositions of the centre of mass of a massive object, they are all, in one way or the other, highly technically challenging.}
These variety of proposals include using supra-molecular complexes \cite{Arndt2014TestingSuperpositions, Fein2019QuantumKDa}, optomechanical systems \cite{scala2013matter,wan2016free,margalit2021realization,wood2022spin,marshman2022constructing,zhou2022mass,Bose1997PreparationMirror, Yin2013LargeCoupling, Lombardo2015, Liao2016, Clarke2018GrowingOptomechanics, Xie2019MacroscopicCoupling,Zheng2020MacroscopicInteraction, Shomroni2020OptomechanicalSubtraction, Zhan2020}, magneto-mechanical systems \cite{Johnsson2016MacroscopicMagnetomechanics, Bose2017, Romero-Isart2017CoherentMicrospheres, Rahman2019LargeNanoparticle, Qin2019ProposalResonators, Pedernales2020MotionalParticles}, and cold atomic systems \cite{Kovachy2015QuantumScale, Pezze2019HeraldedCondensate}. Furthermore, the experimental demonstration of a macroscopic superposition of bulk acoustic mode of matter which has an effective mass of $M\sim 16\,{\rm \mu g}$ \cite{Bild2023SchrodingerOscillator} has been shown towards this end.

{\color{black}In this study, we propose two complementary schemes that have the potential to generate ultra-large spatial superpositions for massive objects, based on the magnetic forces exerted by {\color{black} superconducting flux qubits.}
Briefly, the first scheme is on a levitated magnetic sphere using fixed flux qubits situated near it, and the second scheme is on an isolated floating flux qubit levitated by a fixed magnetic sphere.
{\color{black} These proposals enable the extra-large spatial superpositions, which are not limited by the size of the objects.}
}

\section{Model}
We first describe both our schemes in a general theoretical framework by considering a levitated spherical magnet or a levitated flux qubit.
We assume a coordinate system with the origin at the equilibrium position of this oscillator and consider that an external magnetic actuation to create the superposition state is present along the $z-$ axis.
We denote the mass of the oscillator as $M$ and the oscillation frequency along the $z-$ axis as $\omega_z$.
We consider that the oscillator is in its motional ground state with a ground state width denoted as $\delta z_{\mathrm{zpm}}=\sqrt{\hbar/(2 M\omega_z)}$, where $\hbar$ is the reduced Planck's constant \cite{Cirio2012, Johnsson2016MacroscopicMagnetomechanics}

For the creation of superposition states, we consider logical states $|0\rangle$ and $|1\rangle$ based on superconducting flux qubits. 
The superconducting flux qubit only allows integer numbers of magnetic flux quantum ($\Phi_0= 2.068\times10^{-15}\:{\rm Wb}$) threading the qubit loop and the flux qubit will generate a supercurrent to compensate any non-integer flux quanta threading through its loop. 
Since flux qubits' logical states are determined by the direction of large circulating persistent currents in the superconducting flux qubit  (currents flow either clockwise or counterclockwise) \cite{Cirio2012, Johnsson2016}, the flux qubit can generate logical state-dependent magnetic forces.

{\color{black}We denote the} logical state-dependent magnetic forces generated by flux qubit as \(\Vec{F}^{|q\rangle}_{\mathrm{FQ}}\) for the rest of this letter {\color{black}and we assume that it is along z-axis}, where $|q\rangle$ is the logical state.
The \(\Vec{F}^{|0\rangle}_{\mathrm{FQ}}\) (\(\Vec{F}^{|1\rangle}_{\mathrm{FQ}}\)) is {\color{black}considered to be} along {\color{black} positive (negative) z-axis}, displacing the equilibrium position of the oscillator away from $z=0$, to the new position $z=z_{\mathrm{eq}}$ ($z=-z_{\mathrm{eq}}$).
However, a restoring force, \(\vec{F}_{\mathrm{TR}}=-\) \(M\omega_z^{2} {\color{black}\vec{z}} \) due to the trap of the oscillator along the z-axis acts on the displaced oscillator that pulls it back from both positive and negative z-axis directions towards $z=0$.
As a result the oscillator reaches a new equilibrium position where $\Vec{F}^{|q\rangle}_{\mathrm{FQ}}+\Vec{F}_{\mathrm{TR}}=0$.
When a superposition state $\frac{1}{\sqrt{2}}(|0\rangle+|1\rangle)$ is created using flux qubit (or qubits depending on the scheme) states, the oscillator is pushed to a macroscopic spatial superposition of spatial extent $\Delta z= 2z_{\mathrm{eq}}$.
{\color{black}we express the spatial extent in terms of a dimensionless factor defined as $\chi=\Delta z/\delta z_{\mathrm{zpm}}$.}

We now consider both schemes to estimate $\Vec{F}^{|q\rangle}_{\mathrm{FQ}}$
and thereby estimating \(\chi\) for each case.
In both schemes, we model {\color{black}the uniformly magnetized} spherical magnets as magnetic dipoles.
We model flux qubits as superconducting rings with a radius $R$ and denote the modulus of the supercurrent as $|I|$.
{\color{black}
We consider a flux qubit with loop radius of $R= 183 \:\rm \mu m$ and cross-section radius $a= 5 \:\rm \mu m$, with self-inductance $L= 0.85 \:\rm nH$, for one-flux-quantum jump the current jumps by $\Phi_0 /  L=2.4 \:\rm \mu A$, and for simplicity we  choose $I= \pm 1 \:\rm \mu A$. 
{\color{black} The size of this flux qubit ring is designed to ensure stable trapping, as discussed later. }
}

\subsection{Motional superposition of a levitated spherical magnet}

{\color{black} We focus first on our scheme to generate a macroscopic superposition of a levitated magnet under the influence of a superposition of forces generated by nearby fixed flux qubits. We consider a levitated spherical magnet (for example, Yttrium Iron Garnet (YIG) magnetic micro-particle) with magnetic moment denoted as $\Vec{m}_{0}$.}
{\color{black}The YIG sphere can be magnetically levitated above a superconducting block with a rectangular hole (see [Appendix A] for more details).}

\begin{figure}
\centering
\setlength{\unitlength}{1cm}
\includegraphics[width =0.8\columnwidth]
{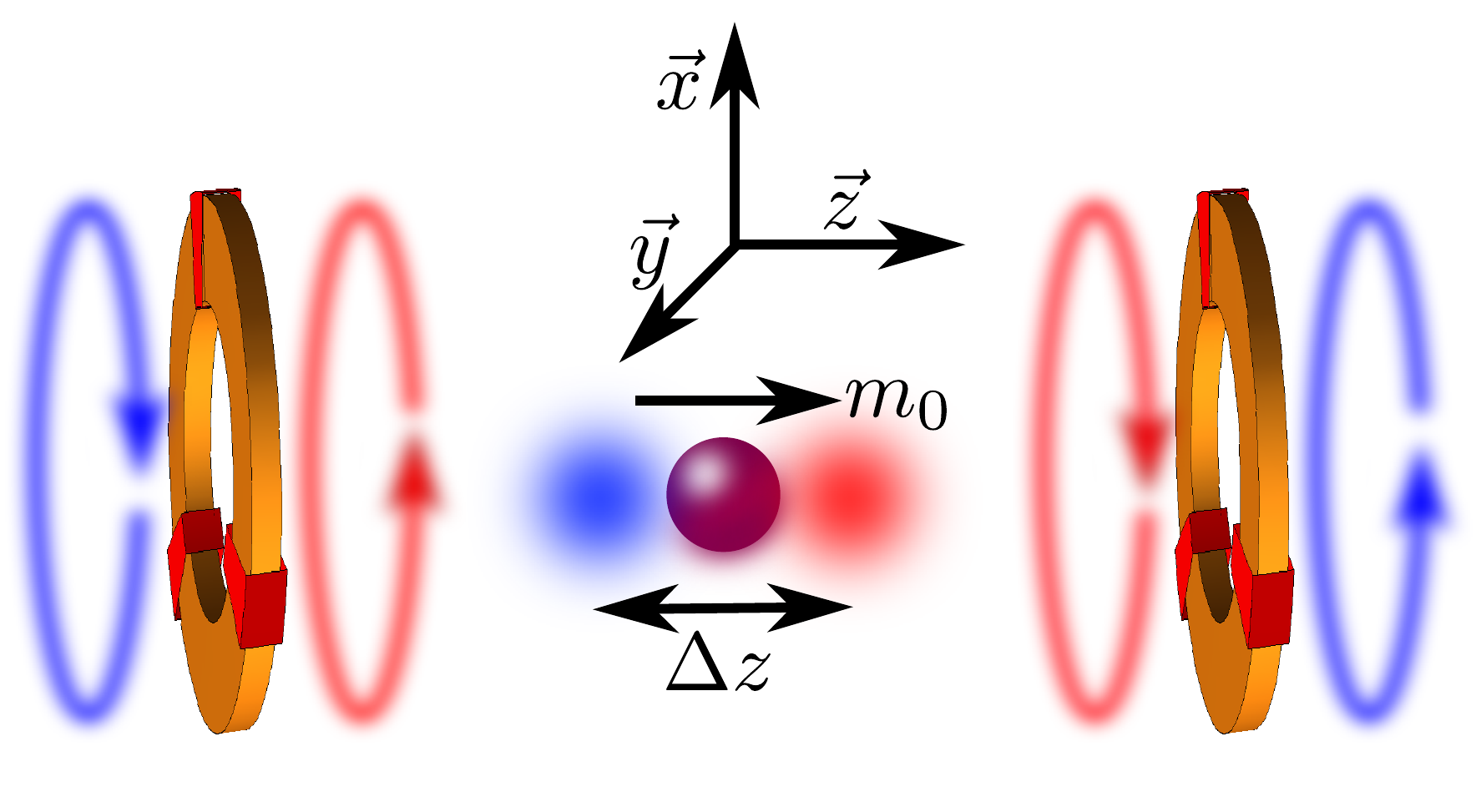}
\caption{Schematic for the creation of superposition states of a levitated magnetic YIG microparticle (oscillator) using superconducting flux qubits (not in scale). 
We depict the oscillator as a sphere that is trapped at the centre.
The nearby flux qubits are shown as rings (orange), containing three Josephson junctions. 
{\color{black} The two flux qubits are connected in the entangled state to make sure that the supercurrent in each qubit has the same magnitude but opposite circulating directions {\color{black} to excite only translational motion}.}
When both qubits are in the $|0\rangle$ $\left(|1\rangle\right)$ state, $I$ flows in opposite circularities depicted in red $\left( \mathrm{blue} \right)$ blurred ring arrows, and the corresponding magnetic field is in anti-Helmholz configuration.
{\color{black} Then the entangled qubit state $|0\rangle|0\rangle$ ($|1\rangle|1\rangle$) shifts the spherical magnet to the right $\left( \mathrm{left} \right)$ by a small distance $z_{\mathrm{eq}}$ as shown by the red $\left( \mathrm{blue} \right)$ blurred sphere. 
By exciting the flux qubits into a superposition state $\sim ( |0\rangle|0\rangle+|1\rangle|1\rangle)$, the micro-particle is driven into a spatial superposition of extent $\Delta z=2 z_{eq}$.}
\label{fig:yig_optical_trap}}
\end{figure}
%

%
We consider {\color{black}the levitated YIG sphere} with the primary trapping axis to be along the $x-$axis (axes depicted in \cref{fig:yig_optical_trap}) and the trapping force is strongest along the $x-$axis compared to the $y-$ and $z-$ axes.
Two entangled flux qubits with identical supercurrents flow with opposite helicity are placed co-axial along the $z-$axis symmetrically on either side of the trapped spherical magnet as shown in \cref{fig:yig_optical_trap}. 
The supercurrents generate magnetic fields emanating from flux qubits in anti-Helmholtz configuration.
When each qubit is in the $|0\rangle$ ($|1\rangle$) state, the \(B_z(z)\) at the midpoint between the two flux qubits vanishes and varies linearly around there, i.e. $B_z(z) \propto z$ ($B_z(z) \propto -z$).
{\color{black}Due to the combination of a vanishing magnetic field and a non-vanishing gradient at the YIG's central location, only translational motion is excited.}
{\color{black}We assume that any change in the magnetic field produced by this YIG at the locations of the flux qubit due to the movement of the YIG within the trap is negligible or compensated appropriately in the experimental implementation (see Appendix B for estimates).}
For this scheme, ${\color{black}F^{|q\rangle}_{\mathrm{FQ}}} = \vec{m}_{0} \cdot (\partial \vec{B}_z/ \partial z)_{|q\rangle}$.

{\color{black}We model the two flux qubits as superconducting rings.
For this we assume that these rings have radii $R$, and each of these flux qubits is symmetrically located at a distance $z^{FQ}_\pm=\pm \eta R$, along the z-axis around the origin, where $\eta$ is a dimensionless scaling factor. 
The flux qubits have supercurrents $I$ circulating in each ring.
We can write down the magnetic field generated by the two superconducting flux qubits at a point on the z-axis $\vec{r}=(0,0,\vec{z})$ as,
\begin{align}
B_{z}(z)_{|q\rangle} = &(-1)^{q} \times \frac{\mu_{0} |I| R^{2}}{2}\nonumber \\ &\left[\frac{1}{((z-\eta R)^{2}+R^{2})^{3/2}} - \frac{1}{((z+\eta R)^{2}+R^{2})^{3/2}}\right],
\label{eq:sm1} 
\end{align}
where $\mu_{0}$ is the vacuum permeability.
The gradient of the magnetic field can be obtained as,
\begin{align}
\left(\frac{\partial B_z} {\partial z}\right)_{|q\rangle} = (-1)^{q} \times \frac{3\mu_{0} |I| R^{2}}{2} \left[-\frac{(z-\eta R)}{((z-\eta R)^{2}+R^{2})^{5/2}} \right. \nonumber\\  \left.+ \frac{(z+\eta R)}{((z+\eta R)^{2}+R^{2})^{5/2}} \right],
\label{eq:sm2}  
\end{align}
Around the origin ($z \rightarrow 0$) the magnetic field gradient is,
\begin{equation}
\left(\frac{\partial B_z} {\partial z}\right)_{|q\rangle} = (-1)^{q}\frac{3\mu_{0} |I|}{R^{2}} \frac{\eta}{(1+\eta^{2})^{5/2}}\;\;,
\label{eq:3}
\end{equation}
which can be approximated to be constant.

We find that for fixed $|I|$ and $R$, \((\partial \vec{B}_z/ \partial z)_{|q\rangle}\) at the origin is maximum when $\eta = 0.5$, that is each of these flux qubits is symmetrically located at a distance $z^{FQ}_\pm=\pm R/2$, along the z-axis around the origin. 
Thus we consider this as the location of the flux qubits for the present scheme.
}

We consider $\Vec{m}_{0}$ to be along {\color{black} the positive z-axis (similar to the experimental study in \cite{Wang2019DynamicsSuperconductor}).}
We can write ${\color{black}|\Vec{m}_{0}|} =~\mu_{0}^{-1}{B_{\mathrm{r}}}V$, where $B_{\mathrm{r}}$ is the {\color{black}remanent magnetic field}.
{\color{black}The new equilibrium position $\pm z_{\mathrm{eq}}$, can be found  by {\color{black} balancing the actuation force from the flux qubits and the restoring trapping force } to find:}

{\color{black}\begin{equation}
\frac{B_{\mathrm{r}}}{\mu_{0}} V  \left(\frac{\partial {B}_z}{\partial z}\right)_{|q\rangle}-\rho V \omega^{2}_{\mathrm{z}} z_{\mathrm{eq}}=0 \;\; .
\label{eq:sm4}
\end{equation}
Using equation (\ref{eq:sm2}), to express the field gradient , we can write equation (\ref{eq:sm4}) as,
\begin{multline}
\frac{3 (-1)^q |I| B_{\mathrm{r}} R^{2}}{\rho \omega^{2}_{\mathrm{z}}}\left[\frac{(z_{\mathrm{eq}}-(R/2))}{((z_{\mathrm{eq}}-(R/2))^{2}+R^{2})^{5/2}} \right. \\ \left. - \frac{(z_{\mathrm{eq}}+(R/2))}{((z_{\mathrm{eq}}+(R/2))^{2}+R^{2})^{5/2}}\right] + z_{\mathrm{eq}}=0\;\;,
\label{eq:sm5}
\end{multline}
where we considered $\eta = 1/2$.
Then for the superposition of $|0\rangle$ and $|1\rangle$, we can obtain spatial superpostion extent as, $\Delta z = 2 z_{\mathrm{eq}}$.
We can write \cref{eq:sm5} using the transformation $z_{\mathrm{eq}} \rightarrow (\Delta z/2)$ to obtain an equation in terms of \(\Delta z\) for numerical calculations as,
}
\begin{multline}
\frac{3 |I| B_{\mathrm{r}} R^{2}}{\rho \omega^{2}_{\mathrm{z}}}\left[\frac{((\Delta z/2)-(R/2))}{((\Delta z/2)-(R/2))^{2}+R^{2})^{5/2}} \right. \\ \left. - \frac{(\Delta z/2)+(R/2))}{((\Delta z/2)+(R/2))^{2}+R^{2})^{5/2}}\right] + (\Delta z/2)=0\;\;,
\label{eq:main1} 
\end{multline}
From equation (\ref{eq:main1}), we can see that \(\Delta z\) {\em is independent of the size of the magnetic sphere} but depends only on $\rho$ and $B_{\mathrm{r}}$. 

\begin{figure}
\centering
\includegraphics[width=\columnwidth]{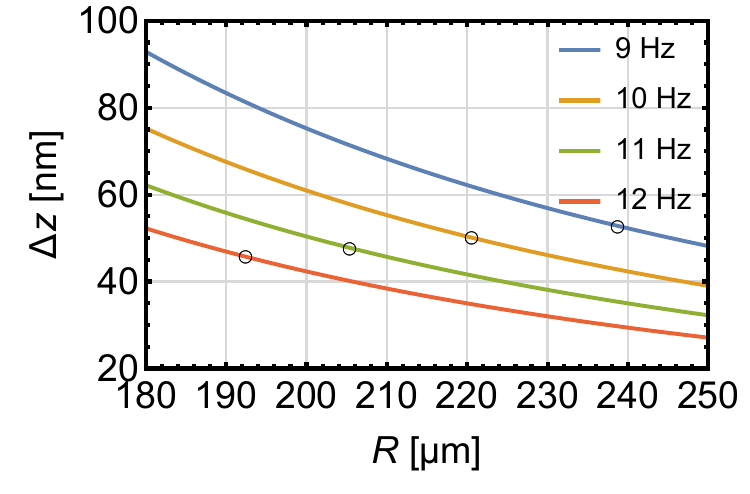}
\caption{
\color{black}{Numerical analyses of spatial superposition separation, $\Delta z$ as a function of the radius of flux qubit, $R$, for trapping frequency $\omega_z/(2\pi)$ as shown in the legends, based on equation (\ref{eq:main1}).} 
{\color{black}The open circle markers on each curve show where $\chi = 10^{6}$ for a YIG sphere with a radius of 25 $\mu \rm m$.
}
}
\label{fig:1}
\end{figure}

We numerically study the extent of the spatial superposition, $\Delta z$ of the YIG sphere as a function of the radius of the flux qubits based on Eq. (\ref{eq:main1}) (please refer Table I in [{\color{black}Appendix F}] for numerical values of parameters used) and the results are shown in \cref{fig:1}.
{\color{black} We consider a YIG sphere of radius 25 $\mu$m, which is  similar in size to other micro-magnets which have been levitated experimentally in magnetic traps \cite{Wang2019DynamicsSuperconductor, Gieseler2020Single-SpinMicromagnets, fuwa2023ferromagnetic}.
For the present scheme, $\Delta z$ can be of the order of hundreds of nanometers and $\chi$ can be of the order of $10^6$.}

\begin{figure}
\centering
\setlength{\unitlength}{1cm}
\includegraphics[width = 0.8\columnwidth]
{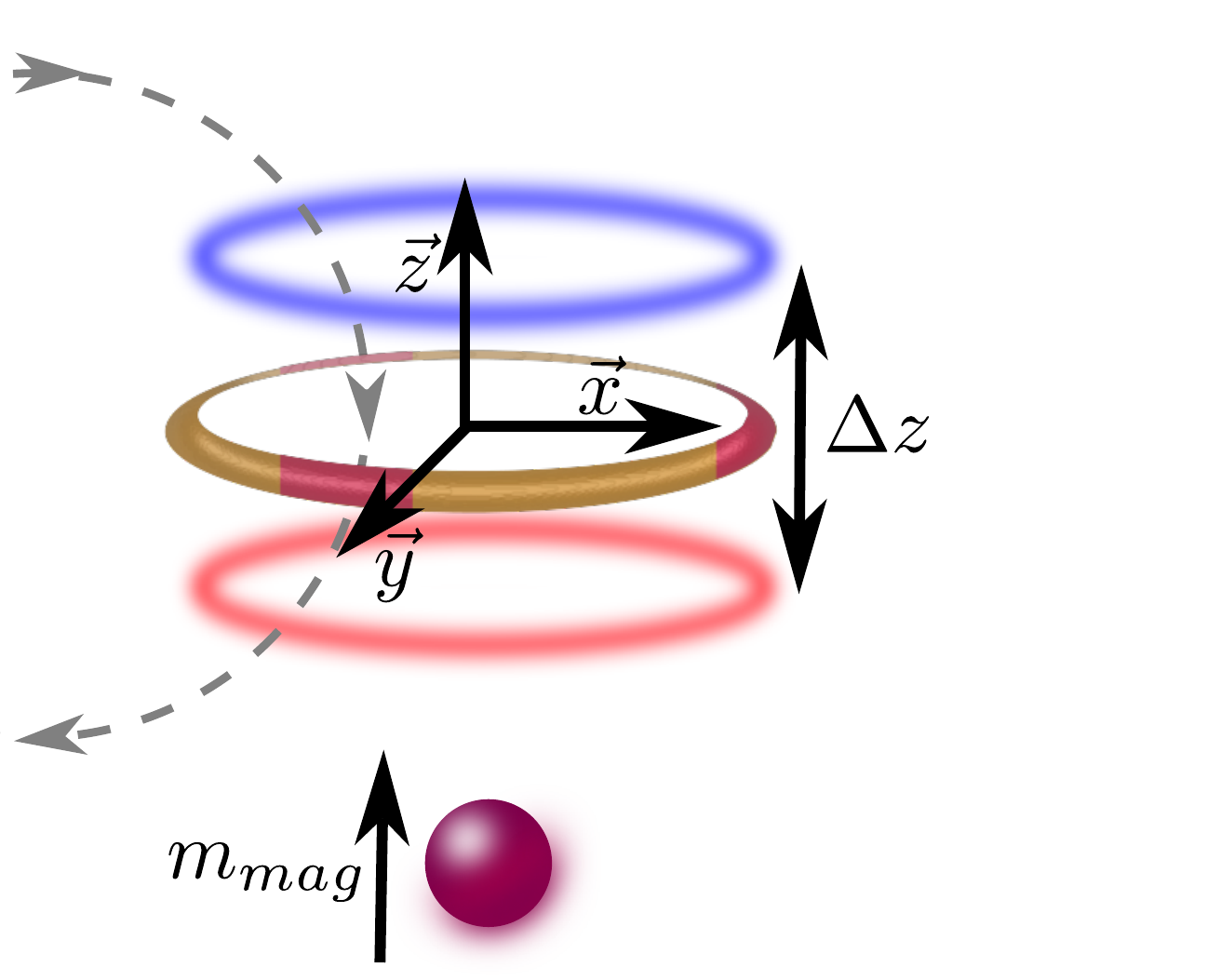}
\caption{Schematic for the creation of macroscopic superposition using the levitated superconducting flux qubit (not in scale). 
The superconducting flux qubit is stably levitated above 
{\color{black} a magnetic dipole (practically a spherical micromagnet) } due to flux-threading conservation and {\color{black}super-diamagnetism}.
The levitated flux qubit is driven inductively using a nearby superconducting circuit. {\color{black}The grey dashed line shows an induced magnetic flux line.}
Levitated flux qubit in $|0\rangle$ ( $|1\rangle$) state displaced upward (downward) as depicted by the blue (red) ring.
By creating a superposition state of $|0\rangle$ and $|1\rangle$ states, the levitated flux qubit is driven into a spatial superposition of extent $\Delta z= 2 z_{eq}$.
}
\label{fig:S2_Setup_schematic}
\end{figure}

\subsection{Motional superposition of a trapped superconducting flux qubit} 

{\color{black} We {now} describe a complementary scheme to generate a macroscopic spatial superposition by levitating an entire superconducting flux qubit ring above a small fixed magnet. The flux qubit contains persistent circulating currents which generate large magnetic moments that, when interacting with the magnetic field produced by the magnet, shift the levitation height of the flux qubit. These circulating currents can be controlled and put into a superposition by inductive driving of the qubit. By driving the qubit into a superposition of circulating currents the qubit is forced into a superposition of heights.} 

We consider the flux qubit (in the shape of a ring) to be magnetically levitated above a spherical magnet, schematically as shown in Fig. \ref{fig:S2_Setup_schematic}.
{\color{black} We model this setup as the levitation of a superconducting ring of finite thickness above a spherical small magnet.} {\color{black} The magnetic field of a uniformly magnetized sphere is exactly that from a magnetic dipole   \cite{Fitzpatrick}}, and the case of an infinitesimally thin superconducting ring levitating above a point magnetic dipole can be analyzed analytically  \cite{Navau2021LevitationMagnetomechanics}.
{\color{black} We assume the origin of the coordinate system is at the center of the ring. The ring is cooled to a superconducting state subject to the zero-field condition (ZFC), and then moved above the magnet with magnetization $m_{mag}\vec{z}$. 
To maintain the flux threading conservation, a current will be induced in the ring with $I=\mu_0 m_{mag} R^2 / (2L(z^2+R^2)^{3/2})$, where $R$ is the radius of the ring, $|z|$ is the distance from the magnet to the center of the ring, and $L=\mu_0 R 
 \text{ln}[8R/r]-2$ is the self-inductance of the ring \cite{Navau2021LevitationMagnetomechanics}. 
The current-carrying ring in the field of the magnet feels a repulsive force which is equal and opposite to the force on the magnet in the field generated by the ring $F_{ring}=({\partial B_z}/{\partial z})m_{mag}$. 
The equilibrium separation between the magnet and the ring $|z|=h$, can be solved by equating the magnetic force $F_{ring}$ to the force of gravity acting on the ring $Mg$ :
\begin{equation}
    \frac{3 \mu_0^2 m_{mag}^2 R^4 z}{4L(z^2+R^2)^4}=Mg\;\;.
    \label{eq:equilibrium}
 \end{equation}
The vertical trap frequency is obtained by $\omega_z=2\pi v_z=\sqrt{k/M}$, where $k=-\partial F_{ring}/\partial z |_{|z|=h}$ is the spring constant, as
\begin{equation}
    v_z=\frac{1}{2\pi}\sqrt{\frac{3 \mu_0^2 m_{mag}^2 R^4(-7h^2+R^2)}{4L(h^2+R^2)^5M}}\;\;.
    \label{eq:frequency}
\end{equation}
We note that in addition to the persistent current generated repulsive force between the ring and the magnet, there is also an additional force due to the superdiamagnetism (Meissner),  of the superconducting ring interacting with the magnet, but the latter is much smaller. This
superdiamagnetism force plays a crucial role in providing the horizontal and tilt stability of the levitated ring (see [Appendix C] for more details).
}

To generate a superposition state, we consider an additional superconducting driver circuit situated nearby the flux qubit {\color{black}without any direct mechanical contact to inductively drive the flux qubit}.
The supercurrent in this driver circuit can generate a magnetic field to change the magnetic flux threading the flux qubit and thereby can inductively drive the flux qubit into the current superposition of $|0\rangle$ or $|1\rangle$ states.
{\color{black} The flux qubit, when in the state $|q\rangle,\,\,q=0,1$, generates a state-dependent magnetic field and field gradient
\begin{equation}
B_{z}(z)_{|q\rangle} = (-1)^{q}  \frac{\mu_{0} |I| R^{2}}{2(z^{2}+R^{2})^{3/2}}\;\;,
\label{eq:B ring} 
\end{equation}

\begin{equation}
\left(\frac{\partial B_z} {\partial z}\right)_{|q\rangle} = (-1)^{q+1}\frac{3\mu_{0} |I| R^2 z}{2(z^2+R^2)^{5/2}} \;\;.
\label{eq:B grdient ring}
\end{equation}
The force on the flux qubit is again equal and opposite to that on the magnet which can be computed as,
\begin{equation}
F^{|q\rangle}_{\mathrm{FQ}}= -\vec{m}_{mag} \cdot (\partial \vec{B}_{z}/\partial z)_{|q\rangle}\;\;.
\end{equation}
At the equilibrium height $|z|=h$, the force on the qubit in state $|q\rangle$, is 
\begin{equation}
F^{|q\rangle}_{\mathrm{FQ}}=(-1)^{q} \frac{3\mu_{0} |I| R^2 h m_{mag}}{2(h^2 + R^2)^{5/2}} \;\;.
\end{equation}
}
Due to this $F^{|q\rangle}_{\mathrm{FQ}}$, the equilibrium height of the flux qubit shifts to a position, $z_\pm=\pm z_{\mathrm{eq}}$, depending on the state of the flux qubit (we choose the $z-$origin to be at the equilibrium height when there is no inductive driving). At these offset equilibria the offset force  $\vec{F}_{\mathrm{FQ}}$ is balanced by the  trapping force $|F_{\mathrm{TR}}| = 2 \pi^{2} r^{2} \rho R \omega_{z}^{2} z_{\mathrm{eq}}$, giving  $\vec{F}^{|q\rangle}_{\mathrm{FQ}} + \vec{F}_{\mathrm{TR}} = 0$. 
We analytically estimate the spatial separation $\Delta z$, between these two equilibria heights as 
\begin{equation}
\Delta z =2z_{eq}\sim \frac{3\mu_{0}m_{mag}|I|Rh}{2\pi^2 r^{2} (h^2+R^2)^{5/2} \rho \omega_z^2}\;\;,
\label{eq:main2}
\end{equation}
when $\Delta z\ll h$.

\begin{figure}[!hbt]
\includegraphics[width=0.9\columnwidth]{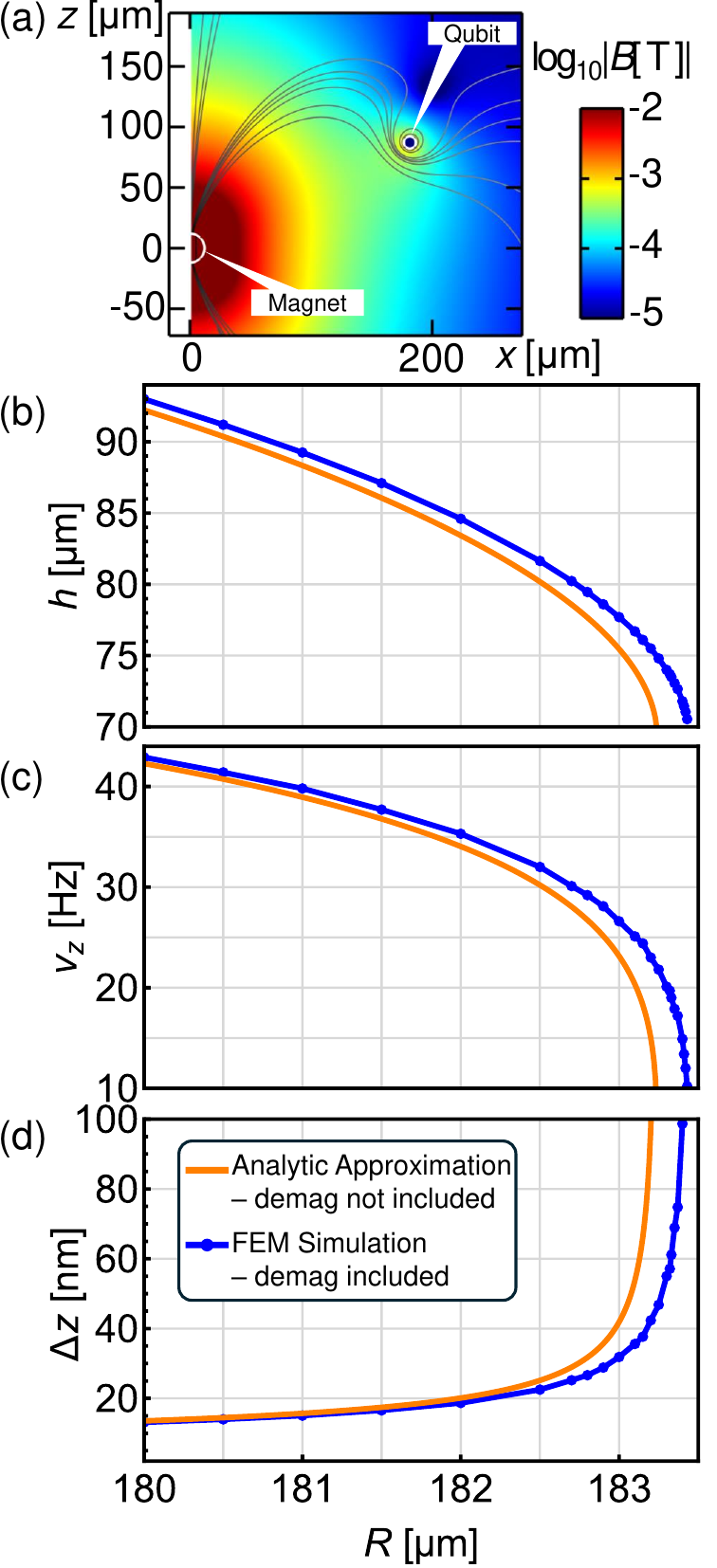}
\caption{
Numerical FEM simulation results of the trapped superconducting flux qubit and its motional superposition. (a) Magnetic flux density $B$ 
{\color{black} in the axisymmetric cross section}
for a magnetic trap where a flux qubit of radius $R=183.4\:\rm \mu m$ is levitated $h=71.8\:\rm \mu m$ above a spherical magnet with a radius $r=12\:\rm \mu m$. We can see the effect of the demagnetizing field of the superconductor on the $B$ field around the levitated flux qubit. {\color{black} (b) Equilibrium position, (c) trap frequency, and (d) superposition separation extent in $z$ direction as a function of $R$.} 
For qualitative comparison, analytical approximations (Eq. \ref{eq:equilibrium} for $h$ in (b), Eq. \ref{eq:frequency} for $v_z$ in (c), and Eq. \ref{eq:main2} for $\Delta z$ in (d)) which do not include the demagnetising effects are plotted as orange curves.
}
\centering
\label{fig:S2_Analysis_And_Simulation_Result}
\end{figure}

The above estimate assumed the flux qubit to be modelled as an infinitesimally thin superconducting ring. To account for a more realistic case where the superconducting ring has finite thickness, we must consider the significant {\color{black}demagnetizing field} generated by the superconductor on the trapping magnetic fields. These demagnetizing fields essentially cause all B-fields within the superconductor to vanish.  To take this effect into account we use a {\color{black}  finite element method (FEM) COMSOL simulation, (blue curves in Fig.\ref{fig:S2_Analysis_And_Simulation_Result})}.
The analytical model in reference \cite{Navau2021LevitationMagnetomechanics}, does not include this demagnetizing effect and their predictions are shown for comparison, {\color{black} (orange curves in Fig.\ref{fig:S2_Analysis_And_Simulation_Result})}.
{\color{black}We are unaware of any published analytical study which includes the demagnetizing field so in [{\color{black}{Appendix D}}], we derive an instructive analytic solution for a simpler example including such demagnetising effects.}
We present the numerical results \cref{fig:S2_Analysis_And_Simulation_Result}, (including demag - blue curves),  which are comparable with the analytic approximations  (no demag - orange curves).
In \cref{fig:S2_Analysis_And_Simulation_Result} (a), we observe a significant perturbation of the fields surrounding the flux-qubit which is expected due to the {\color{black}demagnetizing field} of the superconducting ring and can see that the magnetic field strength at the location of the ring is ($|B|<1 \:{\rm mT}$).
From \cref{fig:S2_Analysis_And_Simulation_Result}(a), we can see that as \(R\) increases, the magnetic field seen by the  flux qubit decreases and we expect the trapping stiffness (and thereby the \(\omega_z\) and \(h\)) to drop as \(R\) increases.
We observe this in \cref{fig:S2_Analysis_And_Simulation_Result}(b) and (c), with a critical value of the radius above which the flux qubit cannot be stably trapped. 
We show the $\Delta z$ in this region, in \cref{fig:S2_Analysis_And_Simulation_Result}(d) and find that $\chi\sim 10^6$ when $R$ $\sim$ $183.4\:\rm \mu m$ (corresponding $\Delta z\sim 100\:{\rm nm}$ and $\nu_z\sim 15\:{\rm Hz}$).
Larger values of $\Delta z$ are predicted from theory but are probably not feasible in experiments.

\section{Decoherence}
{\color{black}  Decoherence is a limiting factor for the generation of a large superposition state in our systems.
Thus, coherence time of the mechanical oscillator and that of the flux qubits are important in our schemes.

The decoherence of mechanical oscillation of the levitated objects are arising from the undesired interactions of the oscillators with the surroundings \cite{Gonzalez-Ballestero2021Levitodynamics:Vacuum}}.
To minimise the interaction of the levitated particle with surrounding that causes mechanical damping we consider the following situations.
In the first scheme, we consider YIG, which is a soft magnet. 
The eddy damping within YIG itself is negligible because it's an electrical insulator \cite{Metselaar1978}.
However, the motion of the oscillator (YIG), may cause eddy damping within the conductors surrounding the oscillator.
We can consider the use of superconducting shields to isolate the trap from other conductors, and it can also screen the external electric and magnetic noise as well.
In the second scheme, the magnet used to lift the ring can be machined using insulating ferrite to avoid eddy damping.
To minimize decoherence due to nearby charge fluctuations, we consider that the levitated objects are electrically neutral. 
This can be achieved using ultraviolet illumination to remove all stray charges on the levitated objects \cite{jin2024towards}.
{\color{black}By theoretically probing the various sources of decoherence in more detail, we identify that gas molecule collision is the dominant damping and decoherence source for our mechanical oscillators (see [Appendix E] for more details).
Here we assume that the levitated system is sufficiently isolated in the dilution fridge to keep the vibration noise level on the oscillators \( \leq 10^{-16} \: g/\rm \sqrt{Hz}\), where $g$ is the gravitational acceleration.
For this, a dilution fridge with pulse tube shutoff should be used to avoid pulse tube noise \cite{uhlig2023dry, franklin2023versatile} and or multi-stages of active and passive vibration isolation systems should be employed to suppress the ground vibration noise \cite{hofer2022high, leng2021mechanical, lee2018vibration, van2023magnetic}.
}
Thus the dominant decoherence rate, which is due to the damping induced by the surrounding gas collisions is then,
\begin{equation}
\Gamma_{\mathrm{g}}~=~\frac{\gamma_{\mathrm{g}}}{\lambda_{db}^2}(\Delta z)^{2}\;\;,
\end{equation}
where \(\gamma_{\mathrm{g}}\) the energy damping rate of the oscillator due to the gas collisions and $\lambda_{db}=\hbar(\sqrt{2 \pi/ M k_B T}$) is the de Broglie wavelength of the oscillator, with $k_{\mathrm{B}}$ is the Boltzmann's constant.
The motional damping rate we define as $\gamma_g$, where \(\gamma_{\mathrm{g}} = \omega_z / Q\), and where \(Q\) is the quality factor of the oscillator.
For the two types of oscillators we are considering, \(Q\) can be generally written as \cite{DeLimaBernardo2013DragGas, Wang2019DynamicsSuperconductor} (see [Appendix E] for more details), 
\begin{equation}
  Q \approx \frac{\pi \rho r \omega_{z}}{6 P_{\mathrm{g}}}  \sqrt{\frac{3 k_{\mathrm{B}} T}{M_{\mathrm{g}}}}\;\;, 
\end{equation}
where $M_{\mathrm{g}}$ is the mass of the gas molecule, and $r$ is the radius (cross-sectional radius) for a spherical (toroid shaped) oscillator.
To minimize  gas collisions  one must operate in the regime of extreme-high vacuum, with pressures $P_g=10^{-16}\:\rm mBar$, which is experimentally feasible using the cyro-pumping effect \cite{hermanspahn2000observation}. To achieve this the vacuum chamber must be operated at  low temperatures $T\sim {\rm mK}$, but such low temperatures are required in anycase for the flux qubit operation.
For a YIG sphere with a radius of 25 $\mu$m, we estimate $Q\sim3.3\times10^{15}$ due to the gas collisions.
Similarly, for the flux qubit ring we estimate {\color{black}the gas collision limited} $Q \sim 3.5\times 10^ {14}$.
The Q-factor, or quality factor, describes the rate of energy loss by the damped oscillator. The rate of loss of the quantum coherence of a large superposition is known as the decoherence rate, and this is usually higher than the rate of loss of energy. This decoherence rate scales quadratically with the size of the superposition $\Delta z$.
Assuming $\Delta z=1\:\rm nm$, we can estimate {\color{black}the decoherence rate due to gas collisions  $\Gamma_{\mathrm{g}}$ for the YIG sphere (flux qubit ring) to be $12 \:\rm Hz$ ($87 \:\rm Hz$) corresponding to a coherence time  $1/\Gamma_{\mathrm{g}}=79 \:\rm ms$ ($11 \:\rm ms$) (see [Appendix E]).}

We consider Type I superconductors (aluminum for flux qubits, and lead for the magnetic trap and shields) for our schemes to avoid flux creep which is an issue in Type II superconductors and should not be present in Type I high purity superconductors.
We assume that the operating temperatures for our schemes are around {\color{black} $10~\:\rm mK$ (can be achieved with a dilution fridge) to ensure that we are in a deeply superconducting regime and to suppress decoherence effects on the flux qubits.}
In our schemes, magnetic field strengths at the location of the flux qubits are well below its critical magnetic field strength of $9.78 \:\rm mT$ \cite{Caplan1965Critical-fieldAluminum}. Aluminum superconducting wire at low temperatures has a critical current density $I_c\sim 10^{10} \:\rm A/m^2$  \cite{kuznetsov2022critical}. {\color{black}For our rings with a cross-section radius of $5\:\rm \mu m$, we verify that the critical current ($\sim 1\:\rm A$), is much higher than any supercurrent considered in the present study, ensuring the superconducting state for the flux qubits.

{\color{black}To achieve $\chi\sim 10^{6},\; (\Delta z \sim 100\:\rm nm)$, using our schemes, an evolution time of half a trap period will be required, which is  $\sim$50~ms (corresponding to very low trap frequencies $\sim$ 10 Hz)}.
{\color{black} To achieve $\chi\sim 10^{4},\; (\Delta z \sim 1\:\rm nm)$, we can increase the trap frequency, and the coherence time required can be reduced to $\sim 100\:\rm \mu s$.
The state-of-the-art flux qubits have coherence times approaching $100\:\rm \mu s$, which can be further improved \cite{yan2016flux,abdurakhimov2019long}.
As we discussed above, the mechanical oscillators with ultra-high motional Q are extremely isolated and can preserve the motional coherence for tens of milliseconds.
Thus, an ultra-large spatial superposition with $\chi\sim 10^{4}$ can be experimentally realized using flux qubits which already exist today.
}

{\color{black} The state purity (e.g. spatial quantum coherence length) provides another quantification of the degree of non-classicality of the decohering quantum superposition. Our estimates of decoherence rates suggest that spatial quantum coherence can be maintained over timescales relevant for the generation of the superposition. 
In the large $t$ limit (for times larger than the localization timescale), the coherence length can be approximated as $\ell(t) \sim 1/\sqrt{2(\gamma_{\mathrm{g}}/\lambda_{db}^2) t}$ \cite{schlosshauer2007quantum}. 
Assuming $\Delta z=1\:\rm nm$, and a decoherence rate $\Gamma \sim 12\:\rm Hz$, one finds $\ell(1 \:\rm ms) \sim 9\: nm$, which is larger than $\Delta z$, indicating that by 1 ms, the spatial quantum coherence of the superposition is still very good (see [Appendix E]). 
}

{\color{black}Realizing a much larger superposition, with $\chi \sim 10^6$, will require a much lower decoherence rate.
The experiment should be implemented at low pressure to reduce the decoherence of the mechanical motion caused by gas collisions. 
Our proposals apply to any flux-based superconducting qubit whose qubit states are defined by large circulating persistent currents. 
The Fluxonium qubit, which has reached millisecond coherence times \cite{somoroff2023millisecond,wang2024achieving}, is a potential candidate.
}

{\color{black}
\section{Preparation and witnessing the superpostion}
To prepare and witness the superpositions we adapt a Ramsey interferometry protocol similar to Ref.\cite{martinetz2020quantum, torovs2021creating}.
We can apply this Ramsey sequence to both schemes: levitated magnet and flux qubit,  to witness the superposition. We assume the resonator is initially in the $|0\rangle$ state and the resonator is in a motional coherent state $|\psi\rangle\sim|\alpha\rangle \bigotimes |0\rangle$, and the Ramsey sequence consists of four steps: 
(1) Apply a $\pi/2$ pulse at time $t_0=0$ to generate a superposition state,
(2) Apply a $\pi$ pulse at time $t_1=\tau$ to flip the qubit state,
(3) Apply another $\pi$ pulse at time $t_2=t_1+\tau^{'}$ to flip the qubit state again,
(4) Apply another $\pi/2$ pulse at time $t_3=t_2+\tau$ and get the final state $|\psi\rangle \sim |\alpha\rangle \bigotimes (\cos{(\frac{\phi_{\rm grav}}{2})}|0\rangle+\sin{(\frac{\phi_{\rm grav}}{2})}|1\rangle) $ to reveal $\phi_{\rm grav}$.
The population of state $|0\rangle$ is $P_0=\cos^2(\frac{\phi_{\rm grav}}{2})$. 
The gravity-induced phase difference $\phi_{\rm grav}=G g \cos{\theta}\sqrt{{2M}/{\hbar\omega_z^3}}\Delta t$, where $g$ is the gravitational acceleration, $\theta$ is the angle between the vertical and $z$ axis, $M$ and $\omega_z$ are the mass and mechanical frequency of the resonator, $\Delta t=2 \tau -\tau^{'}$, and 
{\color{black} $G=m\frac{\partial B}{\partial z} \delta z_{zpm}/(2\pi \hbar),$ is the coupling strength,  where $m$ is the magnetic moment of the magnet, $\frac{\partial B}{\partial z}$ is the gradient of the magnetic field generated by the flux qubit(s), and $\delta z_{zpm}=\sqrt{\hbar/(2 M\omega_z)}$ is the zero point motion extent of the resonator, respectively. We can expand the coupling strength as $G=A {B_r r^3 I} \delta z_{zpm}/(2 \pi R^2 \hbar)$, where $A$ is a dimensionless coefficient, $B_r$ and $r$ are the remanent magnetization and radius of the magnet, $I$ and $R$ are the current and radius of the flux qubit, respectively.
In the first scheme, $A=4\pi\frac{\eta}{(1+\eta^2)^{5/2}}
\approx3.6 $
and in the second scheme, $A=2\pi\frac{\eta}{(1+\eta^2)^{5/2}}
\approx 1.7$. 
Here $\eta$ is the ratio of distance between the spherical magnet and the centre of the flux qubit to the radius of flux qubit ($R$) in both schemes.
}
We obtain coupling strengths ${G}/{2\pi}\approx1.3\times 10^6 \:\rm \mathrm{Hz}$ ($9.4\times 10^6 \:\rm \mathrm{Hz}$) in the first (second) scheme}

\section{Conclusion} 

In summary, we present two complementary schemes to generate massive spatial superpositions on a levitated system based on the superposition of magnetic forces generated by flux qubits. We show that the ratio of the spatial superposition separation to the motional ground state width, $\chi$ can be of the order of $10^4 \text{--} 10^6$. The analyses based on experimental considerations suggest that the experimental realization of our schemes is within reach and thus can pave the way to ultra-sensitive quantum sensing applications.

\section{Acknowledgement}
This research was supported by the Australian Research Council Centre of Excellence for Engineered Quantum Systems (EQUS, CE170100009) and the Okinawa Institute for Science and Technology Graduate University. SB would like to acknowledge EPSRC grant EP/X009467/1 and STFC grant ST/W006227/1.

\section{Appendix A: Magnetic trap for YIG sphere}
The YIG sphere can be levitated above a superconducting block with a rectangular hole (as shown in Figure \ref{fig:trap}). The size and depth of the hole determine the horizontal trap frequencies. The equilibrium height of a magnet above an infinite superconducting plate has been analytically worked out in Ref. \cite{Vinante2020UltralowMicroparticles}. We simulated, using the finite element software COMSOL, the levitation of the YIG sphere above a finite superconducting block with a rectangular indentation as a spatial trap for the YIG. 
As expected, the COMSOL simulation results converges to the analytical predictions based on reference \cite{Vinante2020UltralowMicroparticles}, as the dimensions of the superconductor block becomes larger (see Figure \ref{fig:levheight}).
The simulation also shows that the magnet sphere ($R=25\mu\rm m$), can be stably levitated above a small superconducting block (e.g. $a=b=70~\mu\rm m$), whose size fits between the two flux-based qubits. The Comsol simulation shows that we can engineer the horizontal motional frequency to be extremely low. This levitation setup is small and compact. 

\begin{figure}[hbt!]
    \centering
    \includegraphics[width=0.9\linewidth]{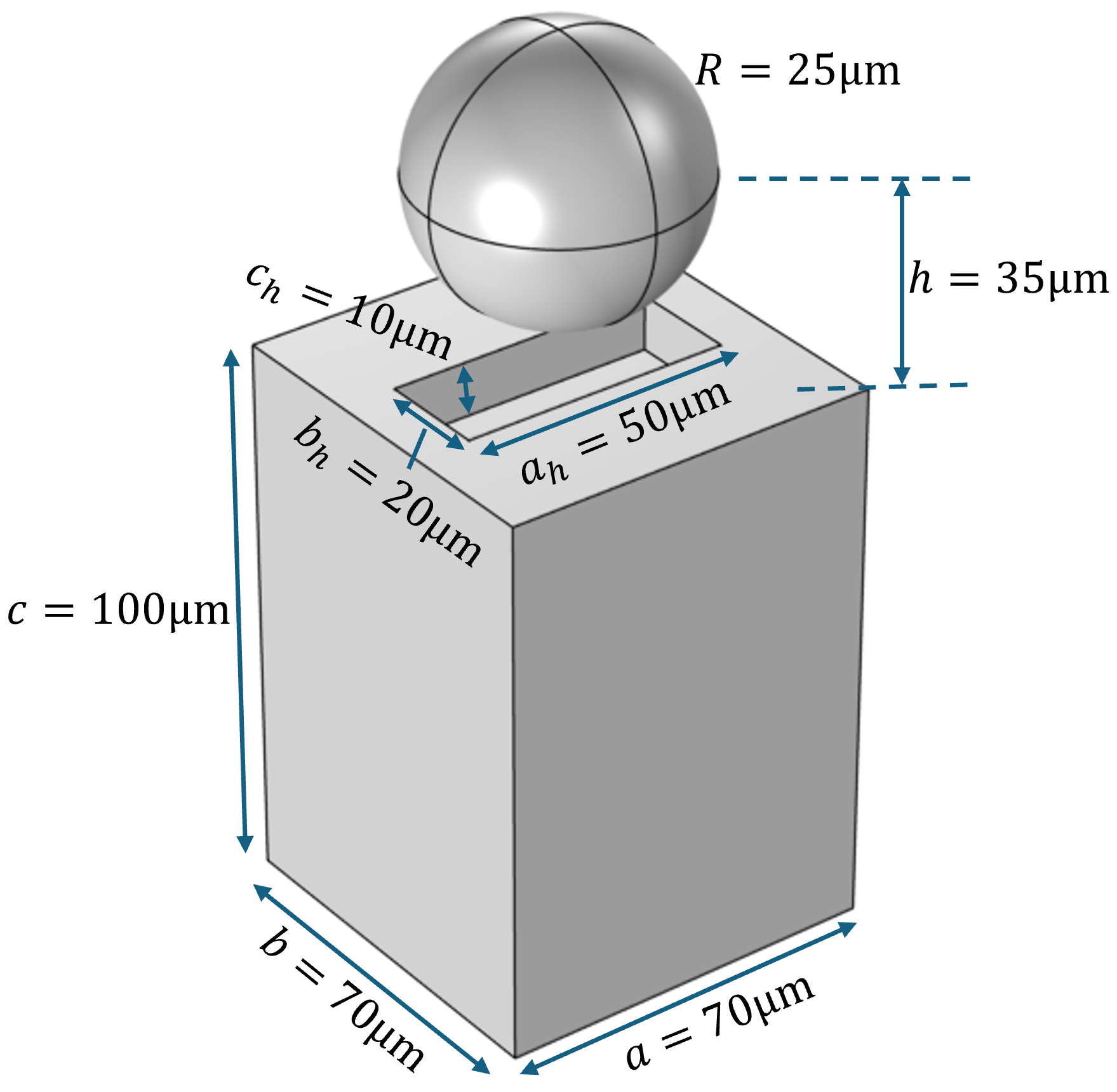}
    \caption{\textcolor{black}{Schematic of the levitated YIG magnet sphere and the superconducting trap with a shallow rectangular hole.
    The magnet sphere with magnetization $B_r=14.32\:\rm mT$ and a radius of $R=25\:\rm \mu m$ is levitated at $h=35\:\rm \mu m$ (the center of the sphere above the top surface of the superconductor block)}.}
    \label{fig:trap}
\end{figure}

\begin{figure}[hbt!]
    \centering
    \includegraphics[width=0.9\linewidth]{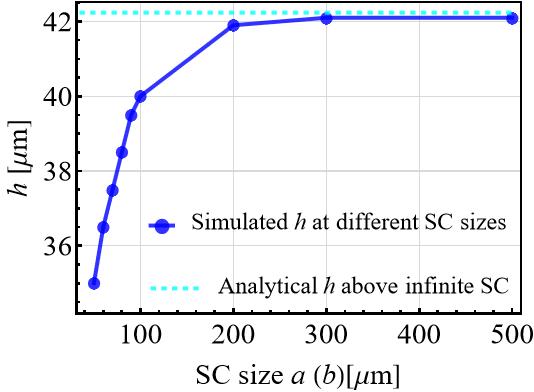}
    \caption{\textcolor{black}{Equilibrium height $h$ of YIG magnet sphere ($R=25\mu\rm m$) above the superconductor (SC) block with different sizes. 
    The dashed cyan line shows $h$ of the magnet above an infinite SC plate, adapted from Ref.\cite{Vinante2020UltralowMicroparticles}. The blue solid line represents the simulated $h$ of the sphere above a SC block with fixed thickness ($c=100\:\rm \mu m$) but different widths (lengths) ($a=b$) in the range of $[50-500]\rm \mu m$}.}
    \label{fig:levheight}
\end{figure}
}

\section{Appendix B: Effect of magnetic field generated by YIG on flux qubit}
\begin{figure}
\centering
{\includegraphics[width = 0.9 \columnwidth]{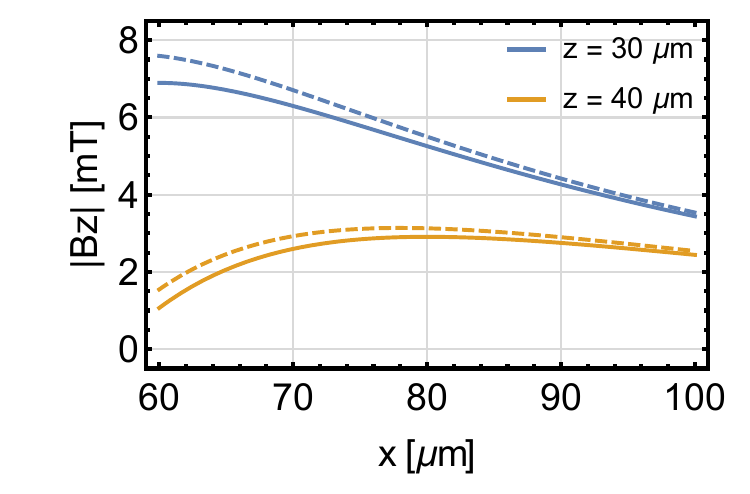}}

\caption{\color{black}{Absolute value of B-field z-component generated by the magnetic sphere} as a function of x position for two different z positions as denoted by the plot legends. The solid curves correspond to the magnetic sphere at the origin and the dashed curves correspond to when the magnetic sphere is shifted to z = 1 $\mu m$}.
\label{fig:Radia_B-field}
\end{figure}

We assumed that the magnetic field produced by the spherical magnet at the location of the superconducting material is below the critical magnetic field of 9.8 mT for Aluminium. To verify this assumption, we consider a spherical magnet of radius 25 $\mu m$ (other parameters are in the \cref{Table1} below), using Radia in Mathematica. {\color{black} Radia is a three-dimensional magnetostatics computer code optimized for solving boundary magnetostatics problems with magnetized and current-carrying volumes using the boundary integral approach} \cite{Chubar1998ADevices}. 
We applied linear anisotropic material for the spherical magnet in the Radia model, with the parallel and perpendicular magnetic field susceptibility of 1 and 0 respectively, with the {\color{black}remanent} magnetic field of YIG along the z-axis.
We simulated the B-field along the x-axis at z = {\color{black}{90}} $\mu m$ and z = {\color{black}{125}} $\mu m$ for a sphere magnet at the origin and also for a sphere located at z = 1 $\mu m$ (Please see Fig. 1 in the main text). 
The results are shown in \cref{fig:Radia_B-field} and from these results, we can see that the assumption is correct for any shifts in position of the YIG within the range presented here.

\section{Appendix C: Horizontal stability of the levitated flux qubit}

We now study the lateral trapping of the levitated flux qubit and will see that it is stably trapped for both small shifts in the lateral position and also for small tilts. We model the levitated flux qubit as a superconducting ring within finite thickness and include the superdiamagnetism and demagnetizing fields, using COMSOL. We find both motions are trapped and stable, confirming the complete trapping of the six degrees of freedom of the levitated flux qubit. We simulate the levitation of superconducting flux qubit using the spherical micro-magnet in a 2D axisymmetric model using the commercial finite element method (FEM) package COMSOL. 
In this FEM model, we consider the flux qubit as a superconducting ring with the ideal Meissner effect. 
We find that the London penetration has a negligible effect on the results due to the large size of the flux qubit.
Furthermore, by performing a fully 3D FEM COMSOL simulation we have determined that the levitated flux qubit is stable both vertically and horizontally, due to the loop shape of the flux qubit. We observe that the magnetic forces and torques will work to restore the qubit's equilibrium position and orientation when flux qubit horizontally shifts or tilts slightly. Thus the trapped flux qubit experiences complete rigid-body trapping.
{\color{black} It's worth noting that Navau et al. expressed concern about the lack of horizontal stability in the case of infinitely thin ring {\cite{Navau2021LevitationMagnetomechanics}}. However, for a real superconducting ring with finite thickness, the Meissner effect of the bulk superconductor will act to keep the ring horizontally stable.}

To test the horizontal stability of the trapped flux qubit, we built a full 3D COMSOL model. In this model, we calculated the x component magnetic force ($F_x$) at the condition of horizontal displacement ($x$). We found that the magnetic force acts as a restoring force when the flux qubit shifts away from the central position in the horizontal direction. Also, when the qubit is tilted, a magnetic torque builds on the flux qubit to restore it to its original horizontal orientation. We predict that the levitated flux qubit will be unstable when the horizontal displacement (or tilting) is too large. We have also calculated the horizontal trap frequency as  $\sim 19\:\rm Hz$, and the tilting frequency as $\sim 9\:\rm Hz$.

\begin{figure}[hbt!]
\centering\includegraphics[width=0.9\columnwidth]{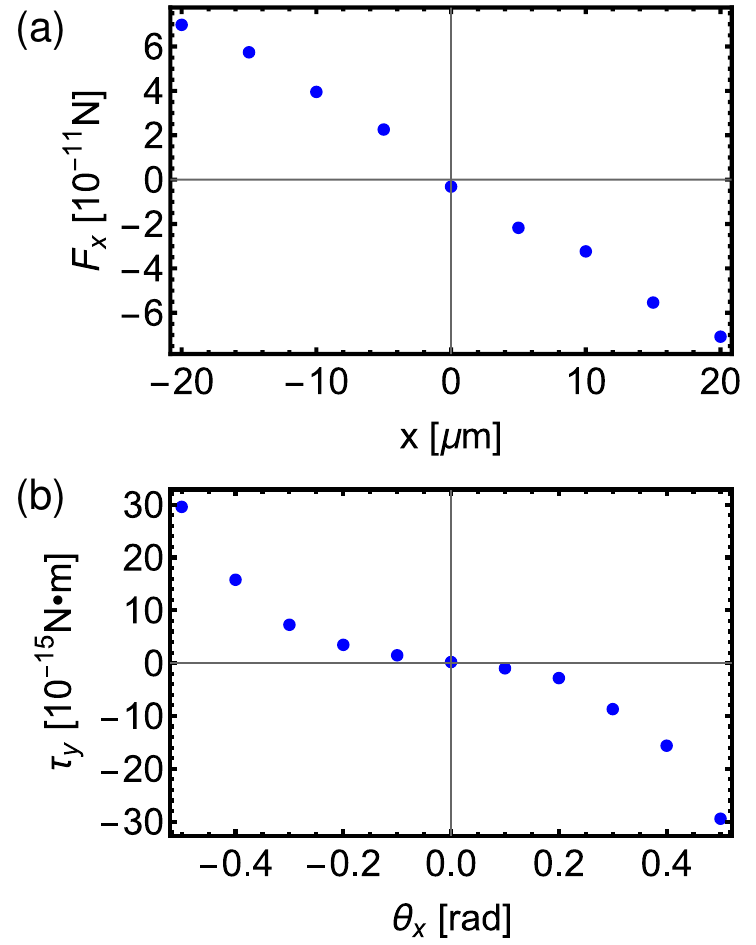}
\caption{The horizontal stability of the levitated superconducting flux qubit. (a) The magnetic force $x$ component $F_{x}$ as a function of the $x$ direction horizontal displacement of the trapped flux qubit. (b) The torque $y$ component $\tau_{y}$ as a function of the tilting angle around $y$. The data was numerically calculated in a full 3D COMSOL model with flux qubit radius $R=183 \:\mu\rm m$.}
\centering
\label{fig:S2S_horizontal_and_tilt}
\end{figure}

\section{Appendix D: Demagnetization effect of superconductors}

{\color{black}To illustrate the effect of {demagnetizing field} and how this can be studied analytically, we consider a simpler example setup: the {diamagnetic} trapping of an SC sphere by two nearby homogenous magnetic spheres (which we consider as magnetic dipoles), arranged in an anti-Helmholtz configuration.
We derive an analytical solution for the trap properties along the trapping axis using an exact method not shown before in the literature, to the best of our knowledge.
We find that the key parameter, the vertical trap stiffness is modified by almost $62.5 \%$ from the case when no {demagnetizing field} is considered and the commercial finite element method (FEM) package COMSOL, provides results that are matching very well with the analytical solution of this example.
}

We show the {\color{black} demagnetizing field} of the superconductor on the magnetic field used to trap it by considering the example of trapping a superconducting (SC) sphere using two dipoles in an anti-Helmholtz magnetic configuration (magnetic moments anti-parallel). This anti-Helmholtz magnetic configuration can be realised experimentally either using current carrying loops as in the section for introducing the first method
or using uniformly magnetised spherical magnets to produce an external magnetic field that is identical to a point magnetic dipole \cite{Edwards2017InteractionsSpheres}.

We first note that a number of authors have considered analytic treatments of the Meissner force of a SC sphere in anti-Helmholtz magnetic fields \cite{Romero-Isart2012,Hofer2019AnalyticField},
 as well as the force between a point dipole and a SC sphere \cite{Coffey2000LevitationSphere,Lin2006AnalyticCylinder}. Many of these works use the image method but there seems to be some concern in the literature regarding this method \cite{Perez-Diaz2007InterpretationLevitation}, and so we do not use the image method in what follows and solve for the magnetic fields and associated forces directly.

\begin{figure}
    \centering
    \includegraphics[scale=0.5]{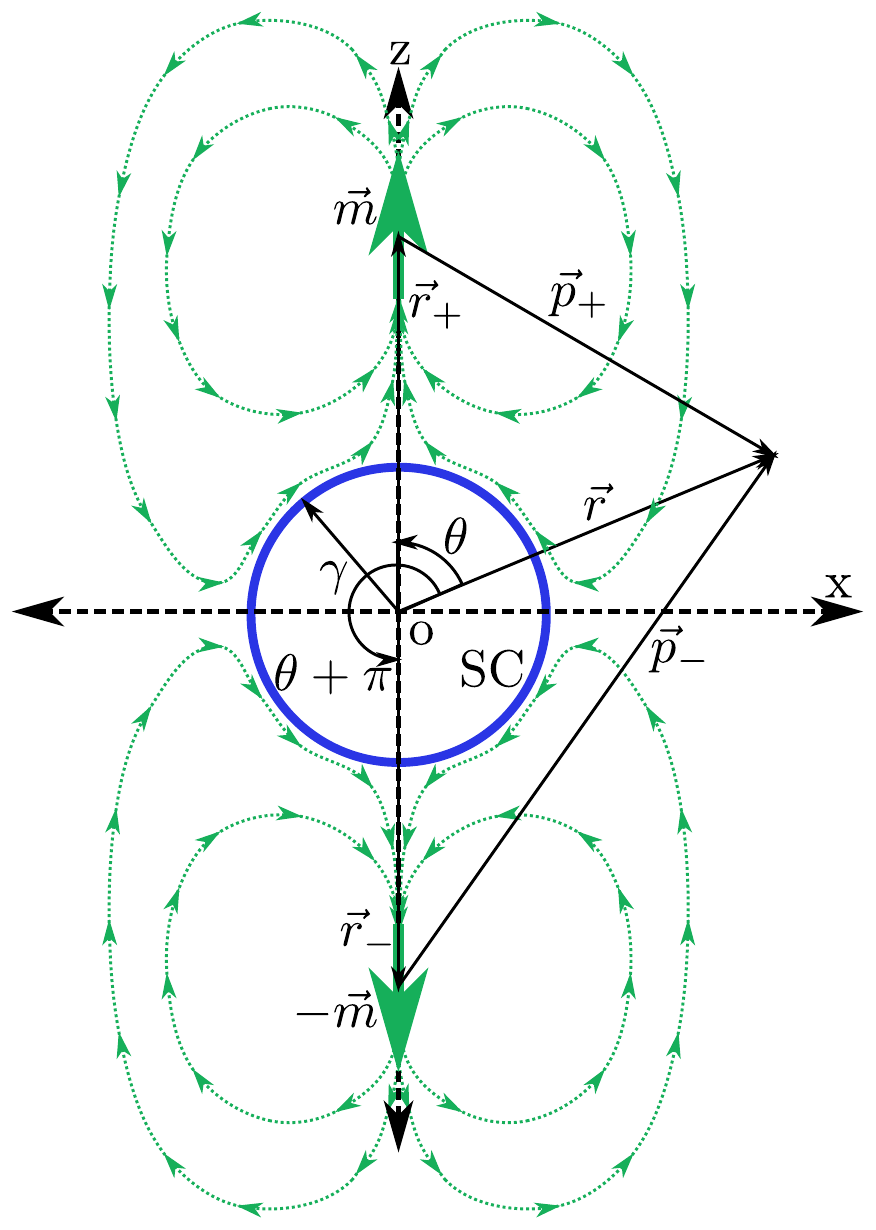}
    \caption{Schematic diagram for the analytical derivation of the magnetic trap properties in z direction. 
    The SC sphere with radius $\gamma$ located at the origin o is shown in blue circle.
    Magnetic dipoles with moment $\vec{m}$ and $-\vec{m}$ are shown in thick green arrows and their field lines are represented in green dotted arrows. 
    $\vec{r}_{+}$ and $\vec{r}_{-}$ are the vectors from origin to magnetic dipoles $\vec{m}$ and $-\vec{m}$ respectively.
    An arbitrary point is chosen for generalized derivation and the $\vec{r}$ is shown from the origin to this arbitrary point.
    $\vec{p}_{+}$ and $\vec{p}_{-}$ are respectively the vectors from magnetic dipoles $\vec{m}$ and $-\vec{m}$ to the arbitrary point of interest.
    }
    \label{fig:AC1}
\end{figure}

We consider the magnets and SC sphere to have centers that all lie on the z-axis, and visualise the geometry in the $x-z$ plane as shown in Figure \ref{fig:AC1}.
%
We analytically derive the induced magnetic field at the SC sphere due to the magnetic field of the two point dipoles, using scalar potentials of the dipoles and enforcing the appropriate boundary conditions at the surface of the SC sphere.
Then we derive an expression for the levitation force by considering the interaction between the resulting magnetic field and the two point dipoles.

We denote $\Phi_{+}$, and $\Phi_{-}$, to be the scalar potentials of the two dipoles located respectively at  $\vec{r}_{+}\sim (0, 0, d_{+})$ and $\vec{r}_{-}\sim (0, 0, -d_{-})$, at either side of the origin along the $z-$axis.
At the arbitrary point $\vec{r}\sim (x, y, z)$, these scalar potentials can be written as ~\cite{Coffey2000LevitationSphere, Coffey2002LondonSphere},
\begin{equation}
\Phi_{\pm}(\vec{r})  = \pm \frac{\mu_{0}}{4\pi}\frac{\vec{m}\cdot\vec{p}_{\pm}}{||\vec{p}_{\pm}||^{3}}\label{eq:AC1},
\end{equation}
where $\mu_{0}$ is the magnetic permeability of free space and $\vec{p}_{\pm}=\vec{r}-\vec{r}_{\pm}$.
In spherical coordinates with $\vec{r}$ radius, $\theta$ as polar angle and $\phi$ azimuthal angle, equations (\ref{eq:AC1}) can be rewritten as,
\begin{equation}
\Phi_{\pm}(\tilde{r}_{\pm}, \theta)  = \frac{\mu_{0} m}{4\pi {d^{2}_{\pm}}} \frac{( (\pm \tilde{r}_{\pm}) \cos{\theta} - 1)}{\Big((\pm {\tilde{r}_{\pm}})^{2}-2 (\pm {\tilde{r}_{\pm}}) \cos{\theta} + 1 \Big)^{\frac{3}{2}}}\label{eq:AC2},
\end{equation}
Here, $m = ||\vec{m}||$, $\tilde{r}_{\pm} = r/d_{\pm}$. There is no $\phi$ dependency due to the axial symmetry of the system.
Expanding using Legendre polynomials, (\ref{eq:AC2}) can be written as,
\begin{equation}
\Phi_{\pm}(\tilde{r}_{\pm}, \theta)  = -\frac{\mu_{0} m}{4\pi d_{\pm}^{2}} \sum_{n=0}^{\infty} (n+1)\left(\pm {\tilde{r}_{\pm}}\right)^{n} P_{n}(\cos\theta)\label{eq:AC3},
\end{equation}
Then the net scalar potential ($\Phi_{\mathrm{m}} \equiv \Phi_{+} + \Phi_{-}$), generated by the two dipoles can be written as,
\begin{multline}
\Phi_{\mathrm{m}}(r, \theta) = -\frac{\mu_{0} m}{4\pi} \sum_{n=0}^{\infty} (n+1) r^{n} P_{n}\left(\cos\theta\right) \\ \left[\left(\frac{1}{d_{+}}\right)^{n+2} + \left(-\frac{1}{d_{-}}\right)^{n+2}\right].
\label{eq:AC4}
\end{multline}
The radial component of the magnetic field generated by the two dipoles is given by, $\vec{B}_{{\mathrm{r}}, {\mathrm{m}}} = -\hat{r} \frac{\partial \Phi_{\mathrm{m}}} {\partial r}$ and in our case we find,
\begin{multline}
\vec{B}_{\mathrm{r}, {\mathrm{m}}}(r, \theta) = \hat{r} \frac{\mu_{0} m}{4\pi} \sum_{n=0}^{\infty} n (n+1) r^{\left(n-1\right)} P_{n}\left(\cos\theta\right) \\ \left[\left(\frac{1}{d_{+}}\right)^{n+2} + \left(-\frac{1}{d_{-}}\right)^{n+2}\right],
\label{eq:AC5}
\end{multline}
We use the subscript r to indicate that we only compute the radial component of $\vec{B}$ and the m subscript indicates that the magnetic field is due to the dipoles and without the SC sphere. 
Introducing the SC sphere results in the introduction of an additional effective magnetic field in response to the field due to the dipoles.
Taken together, the induced and dipole fields must satisfy the Meissner boundary conditions, that the total magnetic field has a vanishing orthogonal component at all superconducting surfaces.

We write down a general solution for the scalar potential corresponding to an induced magnetic field of the SC sphere, due to the Meissner effect as,

\begin{equation}
\Phi_{\mathrm{SC}}(r, \theta) = \sum_{n=0}^{\infty} \frac{C_{n}}{r^{\left(n+1\right)}} P_{n}\left(\cos\theta\right);~ r>\gamma,
\label{eq:AC6}
\end{equation}
where $C_{n}$ is a general coefficient that has to be determined and $\gamma$ is the radius of the SC sphere.
This gives a magnetic field in the radial direction as,
\begin{equation}
\vec{B}_{\mathrm{r}, \mathrm{SC}}(r, \theta) = \hat{r} \sum_{n=0}^{\infty} (n+1) \frac{C_{n}}{r^{\left(n+2\right)}} P_{n}\left(\cos\theta\right).
\label{eq:AC7}
\end{equation}

We now express the total radial component of the magnetic field as, $\vec{B}_{\mathrm{r}, \mathrm{T}}=\vec{B}_{\mathrm{r}, \mathrm{m}}+\vec{B}_{\mathrm{r}, \mathrm{SC}}$.
Considering the perfect Meissner state for the SC sphere, we must have $\vec{B}_{\mathrm{r}, \mathrm{T}}(r=\gamma, \theta) = 0~\forall \theta$. 
This permits us to fix $C_{n}$ to be,
\begin{equation}
C_{n} = - \frac{\mu_{0} m}{4\pi} n \gamma^{\left(2n+1\right)} \left[\left(\frac{1}{d_{+}}\right)^{n+2} + \left(-\frac{1}{d_{-}}\right)^{n+2}\right].
\label{eq:AC8}
\end{equation}
Inserting equation~(\ref{eq:AC8}) into equation~(\ref{eq:AC6}) and equation~(\ref{eq:AC7}), we obtain the scalar potential and the radial component of the induced magnetic field of SC sphere as,
\begin{multline}
\Phi_{\mathrm{SC}}(r, \theta) = - \frac{\mu_{0} m}{4\pi} \sum_{n=0}^{\infty} n \frac{\gamma^{\left(2n+1\right)}}{r^{\left(n+1\right)}} P_{n}\left(\cos\theta\right) \\
\left[\left(\frac{1}{d_{+}}\right)^{n+2} + \left(-\frac{1}{d_{-}}\right)^{n+2}\right],
\label{eq:AC9}
\end{multline}
\begin{multline}
\vec{B}_{\mathrm{r}, \mathrm{SC}}(r, \theta) = -\hat{r} \frac{\mu_{0} m}{4\pi} \sum_{n=0}^{\infty} n (n+1) \frac{\gamma^{\left(2n+1\right)}}{r^{\left(n+2\right)}} P_{n}\left(\cos\theta\right)  \\ \left[\left(\frac{1}{d_{+}}\right)^{n+2} + \left(-\frac{1}{d_{-}}\right)^{n+2}\right].
\label{eq:AC10}
\end{multline}
Choosing $\theta =0$ and $\pi$, the magnetic field at the location of the SC sphere, along the $z$ axis can be obtained from the radial component of the magnetic field. 
In order to compare them with the numerical simulations below, we now convert the magnetic field back to Cartesian coordinates and obtain,
\begin{multline}
\vec{B}^{\pm}_{\mathrm{m}}(z) = \pm \hat{k} \frac{\mu_{0} m}{4\pi} \sum_{n=0}^{\infty} n (n+1) \left(\pm z\right)^{\left(n-1\right)} P_{n}\left(\pm1\right) \\
\left[\left(\frac{1}{d_{+}}\right)^{n+2} + \left(-\frac{1}{d_{-}}\right)^{n+2}\right],
\label{eq:AC11}
\end{multline}
where now $+$ and $-$ in $\pm$ correspond to respectively positive and negative values for the z axis and $\hat{k}$ is the unit vector along the z axis.
Similarly, we find for the induced magnetic field by the SC sphere, with radius $\gamma$ one has,
\begin{multline}
\vec{B}^{\pm}_{\mathrm{SC}} (z) = \mp \hat{k} \frac{\mu_{0} m}{4\pi} \sum_{n=0}^{\infty} n (n+1)\frac{\gamma^{\left(2n+1\right)}}{\left(\pm z\right)^{\left(n+2\right)}} P_{n}\left(\pm1\right) \\
\left[\left(\frac{1}{d_{+}}\right)^{n+2} + \left(-\frac{1}{d_{-}}\right)^{n+2}\right].
\label{eq:AC12}
\end{multline}
From equations (\ref{eq:AC11}) and (\ref{eq:AC12}) we obtain,
\begin{multline}
\vec{B}_{\mathrm{m}}(z) = \vec{B}^{\pm}_{\mathrm{m}}(z) = -\hat{k} \frac{\mu_{0} m}{2\pi} \left[\frac{1}{\left(z-d_{+}\right)^3}+\frac{1}{\left(z+d_{-}\right)^3}\right]
\label{eq:AC13},\\
\end{multline}
\begin{multline}
\vec{B}^{+}_{\mathrm{SC}}(z) = -\vec{B}^{-}_{\mathrm{SC}}(z) = \\
\hat{k} \frac{\mu_{0} m}{2\pi} \gamma^{3}\left[\frac{1}{\left(\gamma^{2}-z d_{+}\right)^3}+\frac{1}{\left(\gamma^{2}+z d_{-}\right)^3}\right]\label{eq:AC14}.
\end{multline}

Then the net magnetic field along the z axis between the two dipoles with the SC sphere is given by the piecewise function:
\begin{equation}
\vec{B}_{\mathrm{T}} (z) = 
\begin{cases}
\vec{B}_{\mathrm{m}}(z) + \vec{B}^{+}_{\mathrm{SC}}(z) & z > \gamma, \\
0 & -\gamma \leq z \leq \gamma,\\
\vec{B}_{\mathrm{m}}(z) + \vec{B}^{-}_{\mathrm{SC}}(z) & z < -\gamma.
\label{eq:AC15}
\end{cases}
\end{equation}
Without the SC sphere present between the two dipoles, $\vec{B}_{T} (z) = \vec{B}_{\mathrm{m}}(z)$.

We compute the z component of the force on the SC sphere and from the dipoles as,
\begin{align}
 \vec{F}_{\mathrm{SC}}(d_{+},d_{-})=&-\hat{k}\frac{\partial } {\partial z}\left(\vec{m}\cdot\vec{B}^{+}_{\mathrm{SC}}(z)\right)\Big|_{z=d_{+}} \nonumber \\ 
 &+\hat{k}\frac{\partial } {\partial z}\left(\vec{m}\cdot\vec{B}^{-}_{\mathrm{SC}}(z)\right)\Big|_{z=d_{-}},
 \label{eq:AC18}
\end{align}
The final form of $\vec{F}_{\mathrm{SC}}$ is given by:
\begin{multline}
    \vec{F}_{\mathrm{SC}} = \hat{k}\frac{3}{2} \frac{\mu_{0} m^{2}}{2\pi} \gamma^{3}\left[\frac{2 d_{-}}{\left(\gamma^{2}-d_{-}^{2}\right)^4}-\frac{2 d_{+}}{\left(\gamma^{2}-d_{+}^{2}\right)^4} \right. \\ \left.
    +\frac{d_{-}}{\left(\gamma^{2}+d_{+}d_{-}\right)^4}-\frac{d_{+}}{\left(\gamma^{2}+d_{+}d_{-}\right)^4}\right]\label{eq:AC19}.
\end{multline}

The properties of the trap can be explored by introducing displacements to the SC sphere.
For this, we consider two parameters the total gap between the centers of two magnets, $d = d_{+}+d_{-}$ and the displacement $2\delta = d_{-}-d_{+}$.
Then with the transformation $d_{\mp} = (d/2)\pm \delta$, we can write the $\vec{F}_{\mathrm{SC}}$ as a function of $\delta$.
We do a series expansion of such a function and from this series that consists only of odd terms we approximately neglect all the higher order terms assuming small displacements, but the first order term.
The resultant $\vec{F}_{\mathrm{SC}}(\delta)$ can be written as,
\begin{equation}
    \vec{F}_{\mathrm{SC}}(\delta) \sim -\hat{k}384\frac{\mu_{0} m^{2}}{\pi} \gamma^{3}\left[\frac{14d^{2}+8\gamma^{2}}{(d^{2}-4\gamma^{2})^{5}}-\frac{1}{(d^{2}+4\gamma^{2})^{4}}\right]\delta \label{eq:AC20}.\\
\end{equation}
For the separation between dipoles is much greater than the diameter of the SC sphere, $d>>2\gamma$, $\vec{F}_{\mathrm{SC}}(\delta)$ becomes,
\begin{equation}
    \vec{F}_{\mathrm{SC}}(\delta) \sim -\hat{k}4992\frac{\mu_{0} m^{2}}{\pi} \frac{\gamma^{3}}{d^{8}} \delta \label{eq:AC21}.\\
\end{equation}
The magnitude of force in (\ref{eq:AC21}) is of the form of $F_{\mathrm{SC}}(\delta) = - k_{\mathrm{z}} \delta$, where $k_{z}~=~4992\mu_{0} m^{2}\gamma^{3}/(\pi d^{8})$ is the spring constant or stiffness of the trap. 
It is also the negative slope of the force-displacement curve that can be drawn from equation (\ref{eq:AC21}).
The gravitational pull also puts an extra force on the SC sphere and displaces it from the origin along the negative z-axis. Then the total force on the SC sphere along the z-axis becomes, $- \hat{k} (k_{\mathrm{z}} \delta +m_{\mathrm{SC}}~g)$, where $m_{\mathrm{SC}}$ and $g$ are the mass of the SC sphere and acceleration due to gravity.
As a result, we can write the equilibrium position as, $\delta_{\mathrm{eq}} \sim ~-m_{\mathrm{SC}}~g/k_{\mathrm{z}}$
If the slope of the force-displacement curve is not going to vary from the origin to the equilibrium position and also approximates the trap as a harmonic one, the angular frequency in radian of the trap along the z axis at $\delta_{eq}$ is given as $\omega_{\mathrm{z}}=\sqrt{k_{\mathrm{z}}/m_{\mathrm{SC}}}$. 

In order to understand how much the {\color{black} demagnetizing field} of the SC sphere affected the dipole magnetic field and the force due to it, we treat the presence of the SC sphere in the magnetic field as a perturbation and do not consider any {\color{black} demagnetizing field} from the SC sphere on the magnetic fields. To identify this case from the previous case we denote the force, stiffness, equilibrium position, and trap frequency, with an over-line above the corresponding symbols. 

The force on the particle along the z-axis is then \cite{OBrien2019, Houlton2018AxisymmetricLevitation}, 
\begin{equation}
 \vec{\overline{F}}_{\mathrm{SC}} =\hat{k}\frac{\chi V}{\mu_{0}} B\frac{\partial B}{\partial z},
 \label{eq:AC22}
\end{equation}
where $B = \sqrt{B_{x}^{2}+B_{y}^{2}+B_{z}^{2}}$, with $B_{x/y/z}$ is the x/y/z component of the magnetic field. Since we are considering a perfect superconductor we can consider $\chi \sim -1$.
In contrast to the previous case, we consider $d_{+} = d_{-}$
In order to obtain  $B_{x/y/z}$, we write down $\Phi_{\mathrm{m}}$ directly in Cartesian coordinates from equation \ref{eq:AC1}  as.
\begin{align}
\Phi_{\mathrm{m}}(x, y, z) = \frac{m}{4\pi}\left(\frac{(z-(d/2))}{(x^{2}+y^{2}+(z-(d/2))^{2})^{3/2}} \right. \nonumber \\ \left.
-\frac{(z+(d/2))}{(x^{2}+y^{2}+(z+(d/2))^{2})^{3/2}}\right).
\label{eq:AC23}
\end{align}
Then we can write $B_{\mathrm{x}}=-\mu_{0}\partial \Phi_{\mathrm{m}}(x, 0, 0)/\partial x$, $B_{\mathrm{y}}=-\mu_{0}\partial \Phi_{\mathrm{m}}(0, y, 0)/\partial y$, and $B_{\mathrm{z}}=-\mu_{0}\partial \Phi_{\mathrm{m}}(0, 0, z)/\partial z$.
The $B_{\mathrm{z}}$ is same as in equation (\ref{eq:AC13}), as $d_{+} = d_{-} = d/2$ 

Since we are interested in the force along the z-axis alone, where $x = y = 0$,
\begin{equation}
 \vec{\overline{F}}_{\mathrm{SC}}(z)=-\hat{k}\frac{V}{\mu_{0}} B_{\mathrm{m}}(z)\frac{\partial B_{\mathrm{m}}(z)}{\partial z},
 \label{eq:AC24}
\end{equation}
Then from equation \ref{eq:AC20} using series expansion and considering only the first order we get,

\begin{equation}
    \vec{\overline{F}}_{\mathrm{SC}} \sim -\hat{k} 3072\frac{\mu_{0} m^{2}}{\pi} \frac{\gamma^{3}}{d^{8}}\delta\label{eq:AC25}.\\
\end{equation}
Comparing equation \ref{eq:AC21} and \ref{eq:AC25}, we can see that the trap stiffness $k_{\mathrm{z}} \sim 1.625~\overline{k}_{\mathrm{z}}$, the equilibrium position $\delta_{\mathrm{eq}} \sim \overline{\delta}_{\mathrm{eq}}/1.625$, and the trap frequency $\omega_{\mathrm{z}} \sim 1.27~\overline{\omega}_{\mathrm{z}}$.

To validate this analytical method, we also built a COMSOL model with the same geometry as that in the above analysis (Figure \ref{fig:AC1}), except for magnet spheres instead of magnetic dipoles. The magnetic field of a magnet sphere with flux density  $Br_{magnet}$ and volume $V_{magnet}$ is identical to that of a magnetic dipole with moment $m=\frac{1}{\mu_{0}}Br_{magnet}V_{magnet}$, so that would not affect the result. In the model, we took the SC sphere radius $\gamma=100\ \mu m$, magnet spheres radius $R_{magnet}=1000\ \mu m$, the distance between two magnets $d=4000\ \mu m$, magnet flux density $Br_{magnet}=1\ T$. In the COMSOL simulation, the {\color{black} demagnetizing field} of the superconductor to the magnetic field used to trap it is well considered. We get the vertical trap 
stiffness $k_{z\_COMSOL}= 0.32\ N/m$, which agrees well with $k_{z}=0.34\ N/m$ from the above {\color{black} demagnetizing field} involved analytical method using the same parameters. Considering $k_{z} \sim 1.625~\overline{k}_{z}$, we can then conclude that the {\color{black} demagnetizing field} of a superconductor in a trapping magnetic field is significant and it would change the vertical trap stiffness by $62.5\%$ in this case.

\section{Appendix E: Decoherence}

In this section, we estimate the decoherence due to unavoidable noises (gas molecule collision, blackbody radiation, and magnetization fluctuation) and introduce measures to avoid other noises.
We assume the proposals work in a vacuum with the pressure of $\sim 10^{-16} \:\rm mBar$ and temperature of $10 \:\rm mK$. Note that these conditions are experimentally practical. $10 \:\rm mK$ temperature is common for a dilution fridge. Due to the cryo-pumping effect, pressure even better than $ 10^{-16} \:\rm mBar$ has been easily reached at $\sim 4 \:\rm K$ in Penning traps \cite{hermanspahn2000observation,gabrielse2006antiproton}.
We find in these conditions the gas molecule collision is the dominant damping source. We estimate the quality factor and decoherence rate of the oscillators which are dominated by the gas molecule collision.  

\subsection{Gas molecule collision}
\textbf{Q factor} \\

The Q factor of a spherical oscillator due to gas molecule collision has been worked out in \cite{Wang2019DynamicsSuperconductor}. 
We adapt it to estimate the Q factors of our spherical YIG oscillator and flux qubit ring. 
We can write down the the Q factor as \cite{Wang2019DynamicsSuperconductor}
\begin{align}
    Q_s&=2\pi\frac{\mathrm{Total~energy~stored}}{\mathrm{Energy~dissipated~per~cycle}}\nonumber\\&\approx2\pi \frac{(1/2)M \omega_z^2 A_m^2}{\bar{F}\times 4 A_m},
    \label{eq_appendix1}
\end{align}
where $M$, $\omega_z$ and $A_m$ are the mass, frequency (along z-axis) and amplitude of the leviated objects respectively.
The $\bar{F}$ is the average drag force on the leviated object (spherical magnet and flux qubit) due to gas collisions \cite{Wang2019DynamicsSuperconductor, DeLimaBernardo2013DragGas}
Assuming that the average velocity ($\approx(1/2)\omega_z A_m$) of the levitated object moving in the gas is much smaller than that of the surrounding gas molecules ($\sqrt{3 k_B T/M_g}$), the average drag force can be written in a simplified form based on \cite{DeLimaBernardo2013DragGas, Wang2019DynamicsSuperconductor} as,
\begin{equation}
    \bar{F}\approx\frac{3}{2}C_{d} A_c P_g \omega_z A_m \sqrt{\frac{M_g}{3k_B T}}\;\; \label{eq_appendix2},
\end{equation}
where $A_c$ is the cross-sectional area perpendicular to the direction of motion, $P_g$ is gas pressure, $M_g$ is the mass of the gas molecule which is assumed as Helium here, $k_B$ is the Boltzmann constant, $T$ is the temperature.
The constant $C_{d}$ is a constant depending on the shape of the levitated object. 
For sphere, $C_{d} = 4/3$ \cite{DeLimaBernardo2013DragGas}.
To calculate $C_{d}$ for the ring, we consider that the ring with a radius of $R$ and cross-section radius of $r$ can be approximated as a cylinder with a radius of $r$ and length $2\pi R$, when the ring moves in the direction perpendicular to the plane of the circle.
For cylinder, $C_{d} = \pi/2$ \cite{DeLimaBernardo2013DragGas}.
Then from equations (\ref{eq_appendix1}) and (\ref{eq_appendix2}), the Q-factor can be written as,
\begin{equation}
    Q=\frac{\pi}{6 } \frac{\omega_z \rho}{P_g} \sqrt{\frac{3k_B T}{M_g}}\left(\frac{V}{C_{d}A_{c}}\right)\;\; ,
\end{equation}
where $\rho$ and $V$ are the density and Volume of the levitated object respectively.
For spherical magnet $A_{c} = \pi r^{2}$, where $r$ is the cross sectional area, and for ring (cylinder), $A_{c} = 2r \times2\pi R$.
Then the parameter $V/(C_{d}A_{c})$ for both sphere and the ring is $r$.
For the first scheme for levitating a spherical magnet, we estimate the Q factor of the $25 \:\rm\mu m$YIG sphere at $10\:\rm Hz$ as $Q_s\sim 3.3\times 10^ {15}$. For the second scheme using an aluminum ring with $r=5 \:\rm\mu m$ oscillating at the direction perpendicular to the ring plane at $10 \:\rm Hz$, the Q factor is estimated to be $Q_r\sim3.5\times 10^{14}$.

\textbf{Decoherence rate} \\
The decoherence rate, which characterizes how long the oscillator's quantum motional state remains coherent, can be estimated from the Q factors
\begin{equation}\Gamma_{\mathrm{g}}=\gamma_{\mathrm{g}}\frac{(\Delta z)^2}{\lambda_{db} ^2} \;\; ,
\end{equation}
where $\gamma_{\mathrm{g}}=\omega_z / Q$ is the damping rate, $\Delta z$ is the extent of superposition state separation, and $\lambda_{db}=\hbar\sqrt{2 \pi/ M k_B T}$ is the de Broglie wavelength of the oscillator, respectively. As we noted in the main text this scales quadratically with the size of the superposition $\Delta z$. Assuming $\Delta z=1\:\rm nm$, we can estimate the decoherence rate for YIG to be $\Gamma_{\mathrm{g}}=12 \:\rm Hz$ corresponding to a coherence time  $1/\Gamma_{\mathrm{g}}=79 \:\rm ms$. For the ring, we get the decoherence rate as $\Gamma_{\mathrm{g}}=87 \:\rm Hz$ with coherence time of $1/\Gamma_{\mathrm{g}}=11 \:\rm ms$.  We will see the decoherence is dominated by the gas molecule collision in our proposed schemes.

\subsection{Blackbody radiation}
We estimate the decoherence caused by blackbody radiation using the YIG sphere. The thermal de Broglie wavelength of massless particles is described by  $\lambda_{bb}=\pi^{2/3}\hbar c /(k_B T)$ \cite{romero2011quantum}. Under the proposed conditions, the thermal wavelength of blackbody radiation $\lambda_{bb}\sim0.5\:\rm m$, is much larger than the superposition separation $\Delta z$. Thus the long wavelength limit can be employed to estimate the Blackbody damping, and we obtain \cite{romero2011quantum} 
\begin{equation}
    \Gamma_{bb}\approx \Lambda_{bb} (\Delta z)^2\;\; ,
\end{equation}
where the decoherence parameters $ \Lambda_{bb} = \Lambda_{bb,sc}+\Lambda_{bb,sb}+\Lambda_{bb,em}$, which 
 consists of three  contributions from scattering, absorption, and emission of thermal photons, are respectively given by
\begin{equation}
    \Lambda_{bb,sc}=\frac{8!\times8\zeta(9) c r^6}{9\pi}\left[\frac{k_B T_e}{\hbar c}\right] ^9 \operatorname{Re}\left[\frac{\epsilon_{bb}-1}{\epsilon_{bb}+2}\right]^2\;\; ,
\end{equation}
\begin{equation}
    \Lambda_{bb,ab}=\frac{16\pi^5 c r^3}{189}\left[\frac{k_B T_e}{\hbar c}\right] ^6 \operatorname{Im}\left[\frac{\epsilon_{bb}-1}{\epsilon_{bb}+2}\right]\;\; ,
\end{equation}
\begin{equation}
    \Lambda_{bb,em}=\frac{16\pi^5 c r^3}{189}\left[\frac{k_B T_i}{\hbar c}\right] ^6 \operatorname{Im}\left[\frac{\epsilon_{bb}-1}{\epsilon_{bb}+2}\right]\;\; ,
\end{equation}
where $\zeta(9)$ is the Riemann $\zeta$ function, $T_i$ is the bulk temperature of the oscillator, $T_e$ is the environment temperature, and $\epsilon_{bb}$ is the 
relative permittivity assumed to be constant over the black body radiation. We take $T_i = T_e =10 \:\rm mK$ and $\epsilon_{bb}=12.5+2i$ \cite{sharma2022low}
, we then estimate the decoherence rate due to blackbody radiation for a sphere with radius $r=25\:\rm \mu m$ with a superposition separation $\Delta z = 1\:\rm nm$ to be $\Gamma_{bb}=5\times10^{-28}\:\rm Hz$. The coherence time $1/\Gamma_{bb}=2\times10^{27} \:\rm s$ is huge and thus the blackbody radiation can be neglected in our proposals.\\

\subsection{Magnetization fluctuation}
The fluctuation of magnetization of the magnets will also lead to the decoherence of the system. 
The power spectrum of the magnetization fluctuation noise is given by \cite{bloch2023scalar}
\begin{equation}
    S_M=\frac{1}{2\pi V}\frac{k_B T}{\omega_M}\mu^{''}(\omega_M)\;\; ,
\end{equation}
where $V$ is the volume of the magnet, $\omega_M$ is frequency and is considered as $1 \:\rm Hz$ here. The magnet has a very low imaginary part of permeability $\mu^{''}(\omega_M)$ if it is fully saturated, and we consider $\mu^{''}(\omega_M)=3\times 10^{-4}\mu_0$ \cite{bloch2023scalar}.  
The noise of the magnetic moment can be expressed as
$ \sqrt{S_m} =\sqrt {S_M} V$.\\

We first consider the decoherence due to the magnetization fluctuation of the YIG magnet.
From equation (\ref{eq:3}) we know the magnetic field gradient at the $z=0$ generated by the flux qubits is
\begin{equation}
    \frac{\partial B}{\partial z}=\frac{3\mu_0 I}{R^2}\frac{\eta}{(1+\eta^2)^{5/2}}\;\; ,
\end{equation}
where $I=1000\:\rm nA$ is the current circulating in the flux qubit, $R=74\:\rm \mu m$ is the radius of the flux qubit, and $\eta=z/R$ is a dimensionless factor describing the distance between flux qubits and the YIG magnet which maximize the gradient field with $\eta=0.5$.
So the magnetic force noise on YIG can be expressed as
\begin{equation}
    \sqrt{S^{mag}_{f}}=\frac{\partial B}{\partial z} \sqrt{S_m}=1.4\times 10^{-28} \:\rm N/\sqrt{Hz} \;\; .
\end{equation}

We consider the force noise due to gas collisions which is given by\cite{torovs2021creating} 
$ \sqrt{ S^{gas}_{f}}=\sqrt{2k_B T M \gamma}=1.3\times10^{-24} \:\rm N/\sqrt{Hz} $.

The force noise due to magnetization fluctuation is much lower than the noise caused by gas collisions. Thus it can be neglected for the decoherence consideration.\\

We now analyze the decoherence on the levitated flux qubit (ring) in the second scheme, caused by the magnetization fluctuation of the fixed permanent magnet.
The levitated ring above a permanent magnet with moment $m_{mag}$  has the magnetic potential  as\cite{Navau2021LevitationMagnetomechanics}
\begin{equation}
    E=\frac{\mu_0^2 m_{mag}^2 R^4}{8L(z^2+R^2)^3} \;\; ,
\end{equation}
where $R\approx183 \:\rm \mu m$ is the radius of the flux qubit, and $L=\mu_0 R [\log(8R/r)-2]$ is the self-inductance of the flux qubit, where $r$ is the radius of the cross-section of the flux qubit. 
The magnetic force is thus given by
\begin{equation}
    f_{mag}=- \frac{\partial E}{\partial z}=
    \frac{3 \mu_0^2 m_{mag}^2 R^4 z}{4L(z^2+R^2)^4} \;\; ,
\end{equation}

So the force noise due to magnetization fluctuation is
\begin{equation}
    \sqrt{S^{mag}_{f}}=\frac{\partial f_{mag}}{\partial m} \sqrt{S_m}= 1.7\times10^{-25} \:\rm N/\sqrt{Hz} \;\; .
\end{equation}

Considering the force noise due to gas collision $ \sqrt{ S^{gas}_{f}}=\sqrt{2k_B T M \gamma}=3.5\times10^{-24} \:\rm N/\sqrt{Hz} $, the noise due to magnetization fluctuation is much smaller, which can also be neglected for the levitated ring.\\

{\color{black}
\subsection{Interaction with thermal phonon bath}

The average phonon occupation number of a mechanical resonator can be written as $\overline{n}=1/(e^{\frac{\hbar \omega_z}{k_B T}}-1)$. The motional thermal bath will interact with the motion of the resonator and change the phonon number. The phonon heating rate, in the high temperature limit, is approximately 
\begin{equation}
\dot{\overline{n}}\approx \frac{k_B T}{\hbar\omega_z} \;\; .
\end{equation}
At a bath temperature of $T=10\:\rm mK$, we 
estimate the phonon heating rate as $\Dot{\overline{n}}\approx 4.0\times 10^{-7} \:\rm Hz$ and $\Dot{\overline{n}}\approx3.7 \times 10^{-6} \:\rm Hz$ for the YIG magnet and the ring, respectively.
}

\subsection{External mechanical vibrations}

\begin{figure}
     \centering
     \includegraphics[width=0.8\linewidth]{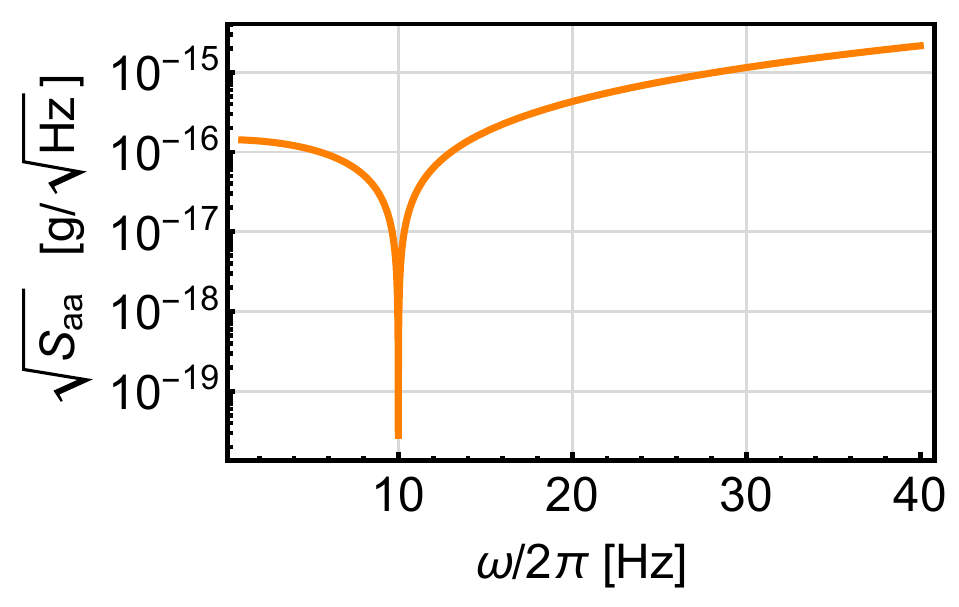}
     \caption{ {\color{black}The vibration acceleration level needed to keep the coherence time longer than $100 \:\rm ms$. Several methods should be taken to isolate the vibration noise below that.}}
     \label{fig:vibration}
 \end{figure}

The external vibrations are detrimental for  superpositions of massive oscillators that possess small ground state extensions.
{\color{black}
The decoherence caused by vibration noise can be estimated as $\Gamma_{vib}\sim M^2 \omega _z^4 S_{zz} {\Delta_z^2}/{(2\hbar^2)}$ \cite{romero2017coherent}, where $M$, $\omega_z$, and $S_{zz}$ are the mass, vertical frequency, and displacement noise PSD of the resonator, respectively.
The acceleration noise PSD can be written as $S_{aa}=S_{zz}((\omega_z^2-\omega^2)^2+(\gamma \omega)^2)$. 
In Figure \ref{fig:vibration}, we plot the vibration acceleration level that should be achieved to keep the coherence time longer than $100 \:\rm ms$. Any vibration source close to the resonance frequency ($10\:\rm Hz$) should be avoided.  Seismic noise, which has low frequency and is one of the main vibration sources, should be suppressed to $\sqrt{S_{aa}}\leq 10^{-16} \: g/\rm \sqrt{Hz}$, where $g$ is the gravitational acceleration.}

{\color{black}
\subsection{Coherence length}
The quality of a macroscopic quantum superposition, particularly one which is subject to decoherence, can be quantified by more quantities that just the spatial separation of the superposition.
The quantum coherence length $\ell$, provides a measure of the quality of the quantum superposition. Under decoherence $\ell$ will decrease in time but if this coherence length remains larger than the spatial separation of the superpositon $\ell>\Delta z$, over a reasonable time duration, then the superposition can be witnessed  via the Ramsey sequence  before the decoherence destroys the superposition.  Our estimates of decoherence rates suggest that coherence can be maintained over the timescales relevant for superposition formation.
We consider the expression of coherence length $\ell$, from Ref. \cite{schlosshauer2007quantum} as
\begin{equation}
    \ell (t)=
    \frac{1}{2} \left[
    \frac{3t^2 + 8\Lambda b^2 t^3 + 12M^2 b^4}
    {2\Lambda t^3+3M^2 b^2 + 4\Lambda^2 b^2 t^4 + 24\Lambda M^2 b^4 t}
    \right]
    ^{1/2}\; ,
\end{equation}
where $b=\Delta z$, $\Lambda = \gamma_g/\lambda_{db}^2$ is the decoherence parameter, and $M$ is mass, respectively. 
In the large $t$ limit ($t$ is much larger than the localization timescale), the coherence length can be simplified as $\ell(t) \sim {1}/{\sqrt{2\Lambda t}}=1/\sqrt{2(\gamma_{\mathrm{g}}/\lambda_{db}^2) t}$. 
Assuming $\Delta z=1\:\rm nm$, decoherence rate $\Gamma = \Lambda (\Delta z) ^2\sim 12\:\rm Hz$, we obtained $\ell(1 \:\rm ms) \sim 9\: nm$, indicating that the quality of the superposition will still be high up to $t=1 {\rm ms}$.
}

\section{Appendix F: Table of variables and their values considered in the present study}

We list all the parameters and values used in Table I and II.
\begin{table*}
\begin{center}
\caption{System parameters and the values used to introduce the first method
\label{Table1}}
\begin{threeparttable}
\begin{tabular}{ p{8cm}p{1.5cm}p{3.5cm}p{1cm}}
\hline\hline
Quantity & Symbol & Value & Unit   \\ \hline
Density of  YIG \cite{Seberson2020SimulationGas}  & $\rho$ & $5110$ &{kg$/m^{3}$}\\
Remanent magnetic field of YIG \cite{Musa2017StructuralSynthesis}  & $B_{r}$ & $14.32 \times 10^{-3}$ & ${\rm T}$ \\
Magnitude of persistant current in SC flux qubit & $I$ & $1000$ & ${\rm nA}$ \\
Mass of the Helium gas molecule & $M_{\mathrm{g}}$ & $6.65 \times 10^{-27}$ & kg \\
Boltzmann constant & $k_{\mathrm{B}}$ & $1.38 \times 10^{-23}$ & $m^{2}$ kg/($K s^{2})$ \\
\hline\hline
\end{tabular}

\end{threeparttable}
\end{center}
\vspace{0mm}
\end{table*}

\begin{table*}
\begin{center}
\caption{System parameters and the values used to introduce the second method
\label{BigTable}}
\begin{threeparttable}
\begin{tabular}{ p{8cm}p{1.5cm}p{3.5cm}p{1cm}}
\hline\hline
Quantity & Symbol & Value & Unit   \\ \hline
Radius of magnet sphere & $a$ & $12$ &${\rm \mu m}$ \\
Radius of SC flux qubit & $R$ & $180-183.4$ & ${\rm \mu m}$ \\
Radius of cross-section of SC flux qubit & $r$ & $5$ & ${\rm \mu m}$ \\
\hline
Magnetic moment of magnet sphere & $m_{mag}$ & $6.912\times10^{-9}$ & ${\rm A\cdot m^{2}}$ \\
Remanent flux density of magnet sphere & $Bz$ & $1.2$ & ${\rm T}$ \\
\hline
Mass of SC flux qubit & $M$ & $(2.41-2.45)\times10^{-10}$ & ${\rm kg}$  \\
Self inductance of SC flux qubit & $L$ & $(8.28-8.48)\times10^{-10}$ & ${\rm H}$ \\ 
\hline
Vacuum magnetic permeability & $\mu_{0}$ & $4\pi\times10^{-7}$ & ${\rm H/m}$ \\
Magnetic flux quantum & $\Phi_{0}$ & $2.068\times10^{-15}$ & ${\rm Wb}$ \\
\hline
Superposition current in SC flux qubit & $I$ & $\pm{1000}$ & ${\rm nA}$ \\
Magnetic moment due to superposition current & $M_{sc}$ & $(1.02-1.06)\times10^{-13}$ & ${\rm A\cdot m^{2}}$ \\
\hline
Equilibrium position of SC flux qubit & $h$ & $70-93$ & ${\rm \mu m}$ \\
Vertical trap frequency & $v_{z}$ & $10-43$ & ${\rm Hz}$ \\
\hline
Superposition states separation & $\Delta z$ & $7-100$ & ${\rm nm}$ \\
Zero point motion amplitude & $\delta z_{zpm}$ & $6\times10^{-14}$ & ${\rm m}$ \\
Superposition separation in terms of $\delta z_{zpm}$ & $\chi$ & $2\times10^{6}$ & NA \\
\hline\hline
\end{tabular}

\end{threeparttable}
\end{center}
\vspace{0mm}
\end{table*}

\bibliography{references}

\end{document}